\documentclass[twocolumn,prb,amsmath,amssymb,amsfonts,superscriptaddress,floatfix,aps,showpacs,notitlepage]{revtex4}
\usepackage{graphicx}
\usepackage{float}
\usepackage{rotating}
\usepackage[utf8]{inputenc} 
\usepackage{natbib}
\usepackage{longtable}
\usepackage{dcolumn}
\usepackage{bm}
\usepackage{color}
\usepackage{xcolor}

\usepackage[caption=false]{subfig}

\usepackage{dcolumn}
\usepackage{bm}
\usepackage[normalem]{ulem}
\usepackage{multirow}

\newcommand{\wideunderline}[2][2em]{%
	\underline{\makebox[\ifdim\width>#1\width\else#1\fi]{#2}}%
}

\newcommand{\be}{\begin{equation}}
\newcommand{\ee}{\end{equation}}
\newcommand{\bea}{\begin{eqnarray}}
\newcommand{\eea}{\end{eqnarray}}

\DeclareUnicodeCharacter{2212}{-}

\begin{document}
	\title{Electron screening and strength of long-range Coulomb interactions in phosphorene: From bulk to nanoribbon}
	
\affiliation{Department of Physics, University of Guilan, 41335-1914, Rasht, Iran}	

\author{Farshad Bagherpour}

\author{Saeed Mahdavifar}

\author{Elham Hosseini Lapasar}

\author{Hanif Hadipour}
\email{hanifhadipour@gmail.com}

	\date{\today}

	
	\begin{abstract}

Experimental observations of anisotropic tightly bound excitons in black phosphorene, and correlated phenomena such as room temperature magnetically active edges in phosphorene nanoribbons (PNRs), sparked discussions on the controversial screening of the Coulomb interaction in phosphorene-based materials. In this way, we investigate the first-principles electronic screening of the long-range Coulomb interaction in black phosphorene from bulk to nanoribbon by employing \emph{ab initio} calculations in conjunction with the constrained random-phase approximation. The bands near Fermi energy (E$_F$) are predominantly $p_z$ orbital characters, and due to the puckering, they are not well separated from the other bands with $s$, $p_x$, and specially $p_y$ characters. This proximity in energy levels increases the contribution of $p_x$/$p_y$ $\rightarrow$ $p_z$ transitions to the polarization function and significantly alters the Coulomb parameters. In semiconducting systems, the on-site Coulomb interaction values (Hubbard $U$) range from 4.1 to 6.5 eV and depend on the correlated subspace, electronic structure, nanoribbon's width, and edge passivation. 
We find an anisotropic behavior of long-range Coulomb interaction along the zigzag and armchair directions in black phosphorene at the static limit. 
Our long-range interaction has revealed a non-conventional screening in semiconducting nanoribbons. We have discovered that screening actually enhances the electron-hole interaction for separations larger than a critical distance $r_c$, which is contrary to what was previously seen in conventional semiconductors. This unique "antiscreening" region helps to explain the large experimentally extrapolated exciton binding energies of PNRs. In unpassivated zigzag nanoribbons, due to a metallic screening channel stemming from quasi-flat edge bands at E$_F$,  we ﬁnd $U/W_b > 1$ (the bandwidth $W_b$) and large gradient of inter-site Coulomb interactions, making them correlated materials. We have investigated the instability of the paramagnetic state of bare ZPNRs toward ferromagnetism using a Stoner model based on the calculated Hubbard $U$ parameters.

	\end{abstract}
	
	
\maketitle

	\section{Introduction}\label{sec1}

Black phosphorus (BP) monolayer, known as black phosphorene, has gained significant attention in recent years due to its unique physical properties ~\cite{Higashitarumizu,Usman, Zhang,Lukasz,Jelver,Biswas,Sui,Bei,Castellanos-Gomez,Li,Liu,Wei}. Bulk BP consists of stacked layers of phosphorene held together by weak Van der Waals forces. Each layer of phosphorene exhibits a puckered honeycomb structure resulting from $sp^3$ hybridization. Black phosphorene is a semiconductor with a direct optical bandgap of approximately 1.7 eV~\cite{Li2}. What sets phosphorene apart from other two-dimensional (2D) materials is its high in-plane anisotropic properties, layer-dependent electronic structure, and high carrier mobility, making it promising for various applications in nano, opto, and thermo-electronics ~\cite{Li2,Wang,Maciej,Xia,Luo,Das1,Zhang1}.
By cutting black phosphorene in specific directions, quasi-one-dimensional phosphorene nanoribbons (PNRs) with armchair and zigzag edges can be produced, referred to as APNRs and ZPNRs, respectively. The properties of PNRs depend on the edge structure, ribbon width, and edge passivation, providing exceptional control over their electronic structure and enabling unique designs for future applications ~\cite{Macdonald,Lee,Feng,Das,Hu,Zhang1}.
Experimental researchers have successfully produced high-quality individual PNRs through techniques such as ionic scissoring of macroscopic BP crystals~\cite{Watts} and electrochemical unzipping of single-crystalline BP into zigzag phosphorene nanobelts~\cite{Liu2} and other recent methods~\cite{Usman,Zhang,Lukasz,Wei,Bei}. All edge hydrogen-passivated PNRs (PNRs:H) exhibit direct semiconductor behavior, with band gaps decreasing as the ribbon width increases. On the other hand, PNRs with unpassivated (bare) edges can be semiconductors in the armchair form but can exhibit magnetic metal behavior in the zigzag form due to the presence of dangling-bond edge states ~\cite{Ramasubramaniam,Ashoka,Han,Li4}.

In low-dimensional materials, reduced screening enhances the Coulomb interaction and affects their transport, magnetic, and optical properties~\cite{Ersoy-1,Wehling,Yekta,Hadipour-2}. This reduced screening has important consequences for the properties of semiconducting PNRs. For example, it leads to the formation of tightly bound excitons~\cite{Shenyang,Yang-1,Choi,Tran,Rodin}. \emph{Ab initio} calculations show that the exciton binding energy of 1 nm wide PNRs is approximately 1.6 eV~\cite{Nourbakhsh}, consistent with strongly localized intrinsic excitons observed at the edge of PNRs in photoluminescence excitation spectroscopy ~\cite{Biswas,Watts,Macdonald}. Such a weak and nonconventional screening of the long-range Coulomb interaction has been reported in other finite-size systems such as graphene nanoribbons (GNRs) ~\cite{Tries,Hadipour,Yang2,Amiri}, MS$_2$ (M=Mo,W)~\cite{Chernikov,Ramezani}, hexagonal boron nitride nanoribbons (\emph{h}-BNNRs)~\cite{Montaghemi}, and semiconducting carbon nanotubes~\cite{Deslippe,Denk}. Furthermore, considering many-body effects within the GW approximation framework, a bandgap correction of approximately 2.3 eV appears in 1.0 nm wide ZPNR:H~\cite{Nourbakhsh}. These significant quasi-particle corrections and the highly anisotropic bright excitons make the study of the role of effective Coulomb interaction in these nanoribbons intriguing.

This is further supported by the fact that in bare PNRs with zigzag edges the room temperature competing FM and AFM ground states have been predicted~\cite{Vahedi,Yang}, which is experimentally evidenced recently via superconducting quantum interference device (SQUID) magnetometry ~\cite{Ashoka}. The presence of strong Peierls instability induced by the well-defined half-filled edge states and correlation effect in ZPNRs makes these systems highly significant~\cite{Du}. Additionally,  bare PNRs with zigzag edges are Mott insulators, which is very rare for $p$ orbital compounds~\cite{Xiaohua}.
Such a rich electronic phase of PNRs, along with other exotic effects such as room-temperature magnetism~\cite{Yang, Ashoka}, topological phase transitions~\cite{Taghizadeh}, and giant spin Seebeck effect~\cite{Farghadan} in ZPNRs, calls for an investigation into the role of Coulomb interactions in these low-dimensional systems. Despite theoretical studies on dielectric screening in few-layer BP~\cite{Roldan}, there has been no \emph{ab initio} study on effective interactions in PNRs so far.

The main goal of this work is the \emph{ab initio} determination of the effective Coulomb interaction between $p$-electrons of the bulk phosphorus, monolayer black phosphorene, and PNRs. These Coulomb parameters can be used in a generalized Hubbard model of phosphorene-based materials to accurately describe the electronic and magnetic phases of PNRs. Overall, the structural puckering of black phosphorene causes $p_x$, $p_y$, ans $s$ states to shift towards the low-energy $p_z$ state. Dielectric screening in black phosphorene shows large anisotropy along different directions, which is consistent with the experimental results~\cite{Wang,Maciej}. We find that the calculated effective Coulomb interactions in quantum-confined PNRs strongly depend on the edge structure, ribbon width, and edge passivation. Furthermore, we find a nonlocal $q$-dependent macroscopic screening in semiconducting PNRs, which are well-screened at short distances and are barely screened or even antiscreened at longer distances. These results are aligned with the observation of localized edge excitonic luminescence in this nanoribbons~\cite{Biswas}.
For completeness, we compared the antiscreening effect in PNRs:H with the corresponding GNRs~\cite{Hadipour} and h-BNNRs~\cite{Montaghemi}.
In unpassivated zigzag nanoribbons, a metallic screening channel originating from quasi-flat edge bands at the Fermi energy (E$_F$) leads to the large gradient of the Coulomb interactions. Due to this localized and short-range interaction, bare ZGNRs can be considered a correlated system.  
The rest of the paper is organized as follows.
The computational methods are presented in Sec.\,\ref{sec2}. Sec.\,\ref{sec3} focuses on the results and discussion, providing a detailed analysis of the effective Coulomb interaction parameters for correlated electrons in single-layer black phosphorene, bulk BP, and PNRs with armchair and zigzag edges. Finally, in Sec.\,\ref{sec4}, the paper is summarized.

\begin{figure}[t]
	\centering
	\includegraphics[width=80mm]{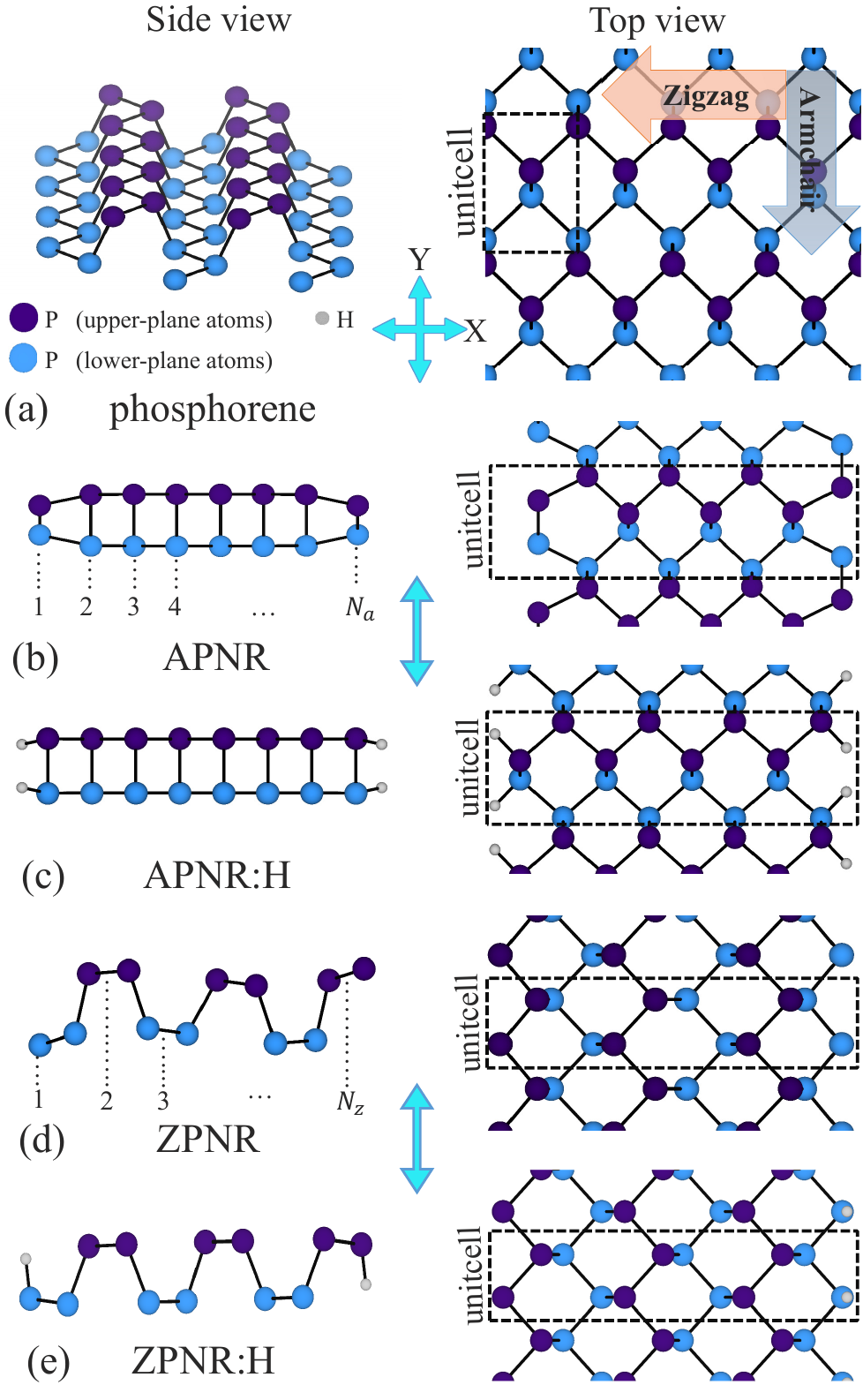}
	\vspace{-0.2 cm}
	\caption{(Color online) Side and top view of the optimized crystal structure of: (a) pristine black phosphorene with armchair and zigzag periodic directions. (b) bare and (c) hydrogen-passivated PNRs with armchair edge. (d) bare and (e) hydrogen-passivated PNRs with zigzag edge. The blue and purple P atoms refer to upper and lower chains and smaller gray balls represent hydrogen atoms.}
	\label{fig:subm1}
\end{figure}

	\section{Computational method}\label{sec2}
	
\subsection{Crystal structure and DFT ground state}
In black phosphorene layers, each phosphorus atom is covalently bonded to three P atoms. Unlike hexagonal planar structures like graphene and \emph{h}-BN, black phosphorene forms a puckered honeycomb structure with $sp^3$ hybridization [Fig.\,\ref{fig:subm1}(a)]. The crystal structure of hydrogen-passivated and bare PNRs is shown in Figs.\,\ref{fig:subm1}(b)-\,\ref{fig:subm1}(e), illustrating both the side and top views.
We designate the armchair and zigzag PNRs as $N_{a}$-APNRs and $N_{z}$-ZPNRs, respectively, following the conventional notation for graphene nanoribbons~\cite{Son}. Here, $N_{a}$ represents the number of dimer lines across the ribbon width, while $N_{z}$ represents the number of zigzag chains [see left side of Figs.\,\ref{fig:subm1}(b) and \,\ref{fig:subm1}(d)]. For APNRs:H and ZPNRs:H, we consider $3\leq N_{a}\leq10$ and $2\leq N_{z}\leq8$, respectively. In the case of bare PNRs, we choose $3\leq N_{a}\leq8$ for APNRs and $N_z=7$ for ZPNRs. To simulate both groups of unit cells and prevent interactions between periodic images, we use a vacuum space of 20 {\AA} in both the in-plane and out-plane directions of the ribbons. All atomic positions are fully relaxed until the force on all atoms becomes less than 0.01 eV/{\AA}.

 The ground-state DFT calculations within the full-potential linearized augmented plane wave (FLAPW) method are carried out with the FLEUR code~\cite{fleur} using a generalized gradient approximation (GGA) parameterized by Perdew, Burke, and Ernzerhof (PBE)~\cite{Perdew} for the exchange-correlation energy functional. To determine the ground state, We used $22\times1\times1$ and $22\times22\times1$ ($12\times12\times12$) $k$-point grid for unit cells of PNRs and pristine black phosphorene (bulk BP), respectively. Also, for all systems, we applied $k_{max}$ =4.5 bohrs$^{-1}$ cutoff for the wave functions. 
 The maximally localized Wannier functions (MLWFs)
are constructed using the WANNIER90 library~\cite{Pizzi,Marzari,Mostofi,Freimuth}.

\subsection{cRPA method}

 The partially (fully) screened Coulomb matrix elements are calculated in the cRPA (RPA) method with the SPEX code (for more details see Refs.~\cite{cRPA_1,cRPA_2,cRPA_3,SPEX}). So, in the following, we provide a brief description of cRPA method.

 The fully screened Coulomb interaction is defined as
\begin{equation}
W(\boldsymbol{r},\boldsymbol{r}',\omega)=\int d\boldsymbol{r}''  \epsilon^{-1}(\boldsymbol{r},\boldsymbol{r}'',\omega) V(\boldsymbol{r}'',\boldsymbol{r}'),
\label{fullysw}
\end{equation}

where $\epsilon(\boldsymbol{r},\boldsymbol{r}'',\omega)$ is the dielectric function and $V(\boldsymbol{r}'',\boldsymbol{r}')$ is the bare Coulomb interaction. In the RPA framework, dielectric function is approximated by

\begin{equation}
\epsilon(\boldsymbol{r},\boldsymbol{r}',\omega)=\delta(\boldsymbol{r}-\boldsymbol{r}')-\int d\boldsymbol{r}'' V(\boldsymbol{r},\boldsymbol{r}'')P(\boldsymbol{r}'',\boldsymbol{r}',\omega),
\label{rpadiel1}
\end{equation}
where the polarization function $P(\boldsymbol{r}'',\boldsymbol{r}',\omega)$ is given by
\begin{equation}
\begin{gathered}
P(\boldsymbol{r},\boldsymbol{r}',\omega)=\\
\sum_{\sigma} \sum_{\boldsymbol{k},m}^{occ} \sum_{\boldsymbol{k}',m'}^{unocc} \varphi_{\boldsymbol{k}m}^{\sigma}(\boldsymbol{r}) \varphi_{\boldsymbol{k}'m'}^{\sigma*}(\boldsymbol{r}) \varphi_{\boldsymbol{k}m}^{\sigma*}(\boldsymbol{r}') \varphi_{\boldsymbol{k}'m'}^{\sigma}(\boldsymbol{r}')  \\
\times\Bigg[ \frac{1}{\omega-\epsilon_{\boldsymbol{k}'m'}^{\sigma}-\epsilon_{\boldsymbol{k}m}^{\sigma}-i\eta} - \frac{1}{\omega+\epsilon_{\boldsymbol{k}'m'}^{\sigma}-\epsilon_{\boldsymbol{k}m}^{\sigma}-i\eta} \Bigg].
\end{gathered}
\label{rpapol1}
\end{equation}
\newline

where $\varphi_{\boldsymbol{k}m}^{\sigma}(\boldsymbol{r})$, $\epsilon_{\boldsymbol{k}m}^{\sigma}$ are chosen to be the Kohn-Sham eigenfunctions and eigenvalues obtained from DFT calculation with spin $\sigma$, wave number $\boldsymbol{k}$, and band index $m$. $\eta$ is a positive infinitesimal and the tags $occ$ and $unocc$ above the summation symbol means that the only terms involving the products of occupied and unoccupied states remain.

\begin{figure}[t]
	\centering
	\includegraphics[width=84mm]{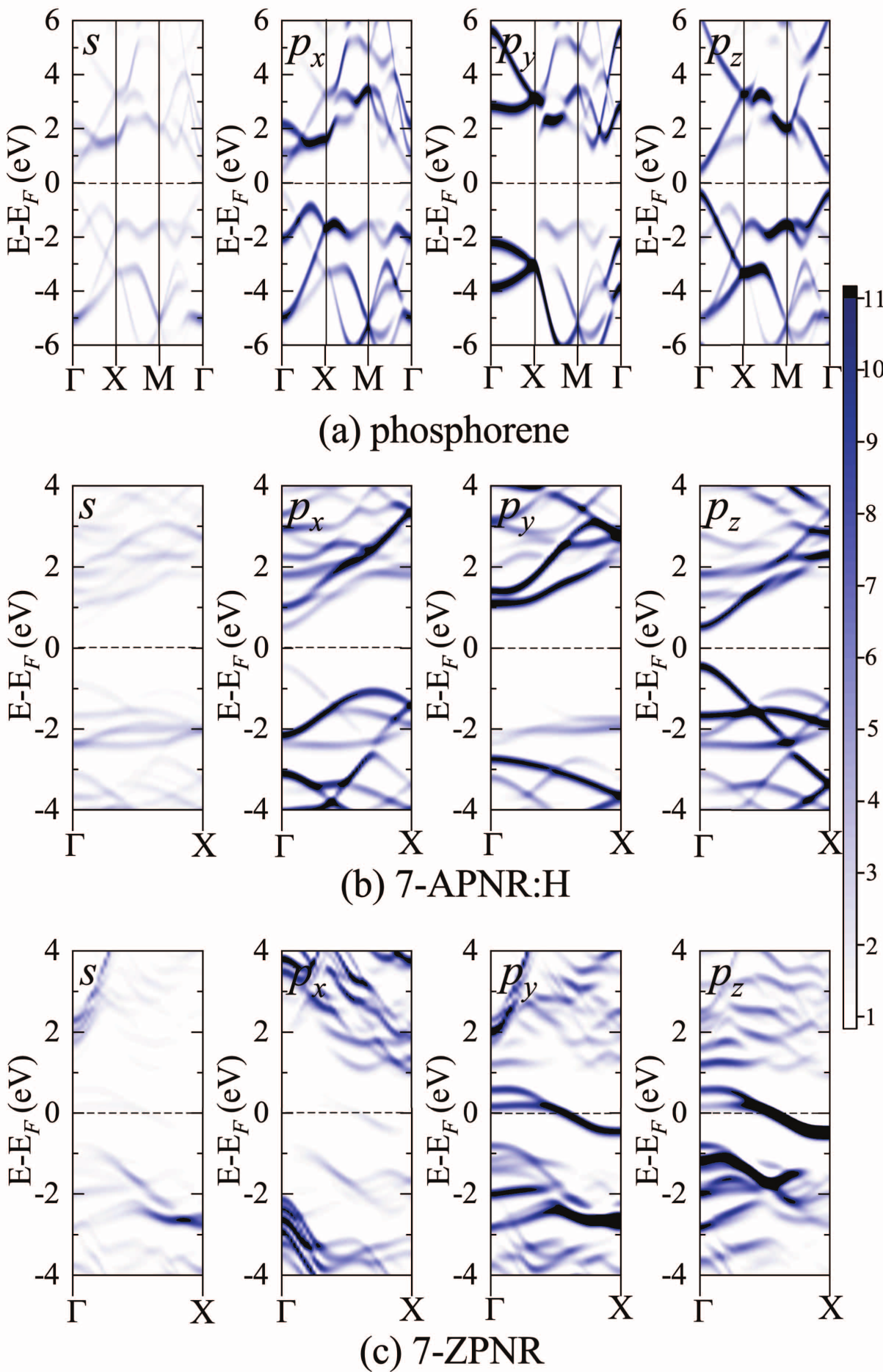}
	\vspace{-0.2 cm}
	\caption{(Color online) Orbital-projected band structure of (a) black phosphorene, (b) 7-APNR:H, and (c) 7-ZPNR. The Fermi level is set to zero energy.}
	\label{fig:subm2}
\end{figure}

In the cRPA approach, we are able to calculate the partially screened Coulomb interaction between the localized electron. To exclude the screening due to the correlated subspace, we divide the full polarization function to the two parts: $P=P_{l}+P_{r}$, where $P_{l}$ includes only transitions between the localized $p$ states and $P_{r}$ is the remainder. 
To recognize which orbital should be considered as $P_{l}$ correlated subspace and reveal the mixing of $p_{z}$ states with other bands, the projected band structures for black phosphorene, 7-APNR:H and,  bare 7-ZPNR are presented in Fig.\,\ref{fig:subm2}. For all systems, the contribution of $s$ orbitals states around E$_{F}$ is almost negligible compared to $p_{x}$, $p_{y}$, and $p_{z}$ orbitals. 

Then, the frequency-dependent partially Coulomb interaction (Hubbard $U$) is given by 

\begin{equation}
U(\omega) = [1-VP_{r}(\omega)]^{-1}V.
\end{equation}

\begin{table*}[!ht]
	\caption{On-site $U_{00}$, nearest-neighbor $U_{01}$, next nearest-nieghbor $U_{02}$, and third nearest-nieghbor $U_{03}$  of black phosphorene and bulk BP for $p$ and $p_{z}$ correlated subspace. The results for graphene are also given for comparison. The bare $V$, partially screened $U$ (cRPA), and fully screened parameters $W$ (RPA)  are defined in computational method section. The distance between two P atoms is shown by $d$ in {\AA}.}
	\centering
	\begin{ruledtabular}
		\begin{tabular}{ccccccccccccccccccccccc}
			\rule{0pt}{4mm}%
& \multicolumn{4}{c}{black phosphorene ($p$)} &&\multicolumn{4}{c}{black phosphorene ($p_{z}$)} && \multicolumn{4}{c}{Bulk BP ($p$)} &  &\multicolumn{3}{c}{Graphene ($p_{z}$)}
\rule{0pt}{4mm}%
\\ \cline{2-5} \cline{7-10} \cline{12-15} \cline{17-19}
$U_{0,j}$(eV) &V bare&cRPA&RPA&$d$ (\AA)&&V bare&cRPA&RPA&$d$ (\AA)&&V bare&cRPA&RPA&$d$ (\AA)&&V bare&cRPA&RPA
\rule{0pt}{4mm}%
\\ \hline
\rule{0pt}{4mm}%
$U_{00}$&13.23&5.51&3.81&0.00&&13.24&4.09&3.79&0.00&&15.42&4.99&3.73&0.00&&16.7 \cite{Ersoy}&8.7\cite{Ersoy}&4.5\cite{Montaghemi}\\
$U_{01}$&6.03 &2.36&1.45&2.22&&5.74 &1.47&1.44&2.22&& 6.26&1.25&0.78&2.25&& 8.5 \cite{Ersoy}&4.0\cite{Ersoy}&1.5\cite{Montaghemi}\\
$U_{02}$&6.04 &2.35&1.42&2.24&&6.72 &1.93&1.42&2.24&& 4.25&0.65&0.37&3.34&& 5.4 \cite{Ersoy}&2.5\cite{Ersoy}&0.9\cite{Montaghemi}\\
$U_{03}$&4.20 &1.67&1.02&3.35&&4.03 &1.10&1.02&3.35&& 2.37&0.33&0.19&3.41&& 4.7 \cite{Ersoy}&2.2\cite{Ersoy}&0.5\cite{Montaghemi}\\
		\end{tabular}
		\label{table1}
	\end{ruledtabular}
\end{table*}

Eventually, using MLWFs as a basis, the effective Coulomb interaction $U$ is calculated by
	
 \begin{equation}
	\begin{gathered}
	U_{in_{1},jn_{3},in_{2},jn_{4}}^{\sigma_{1},\sigma_{2}}(\omega)= \\
	\int \int d\boldsymbol{r}d\boldsymbol{r}' w_{in_{1}}^{\sigma_{1}*}(\boldsymbol{r}) w_{jn_{3}}^{\sigma_{2}*}(\boldsymbol{r}') W(\boldsymbol{r},\boldsymbol{r}',\omega) w_{jn_{4}}^{\sigma_{2}}(\boldsymbol{r}') w_{in_{2}}^{\sigma_{1}}(\boldsymbol{r}).
	\end{gathered}
	\label{hubudef31}
  \end{equation}

where,  $w_{in}^{\sigma}(r)$ is a MLWF at site $R$ with orbital index $n$ and spin $\sigma$. If we only consider the static limit ($\omega=0$), the average on-site Coulomb matrix elements are estimated as  \begin{equation}
	U = 1/L \sum_{n}U_{Rnn:nn}^{\sigma_{1},\sigma_{2}} (\omega=0),
	\end{equation}
and inter-site (long-range) elements between two sites are defined as
    \begin{equation}
     U(R-R^{\prime}) = 1/L \sum_{n}U_{Rnn:R^{\prime}nn}^{\sigma_{1},\sigma_{2}} (\omega=0),
	\end{equation}

\begin{figure}[b]
	\centering
	\includegraphics[width=80mm]{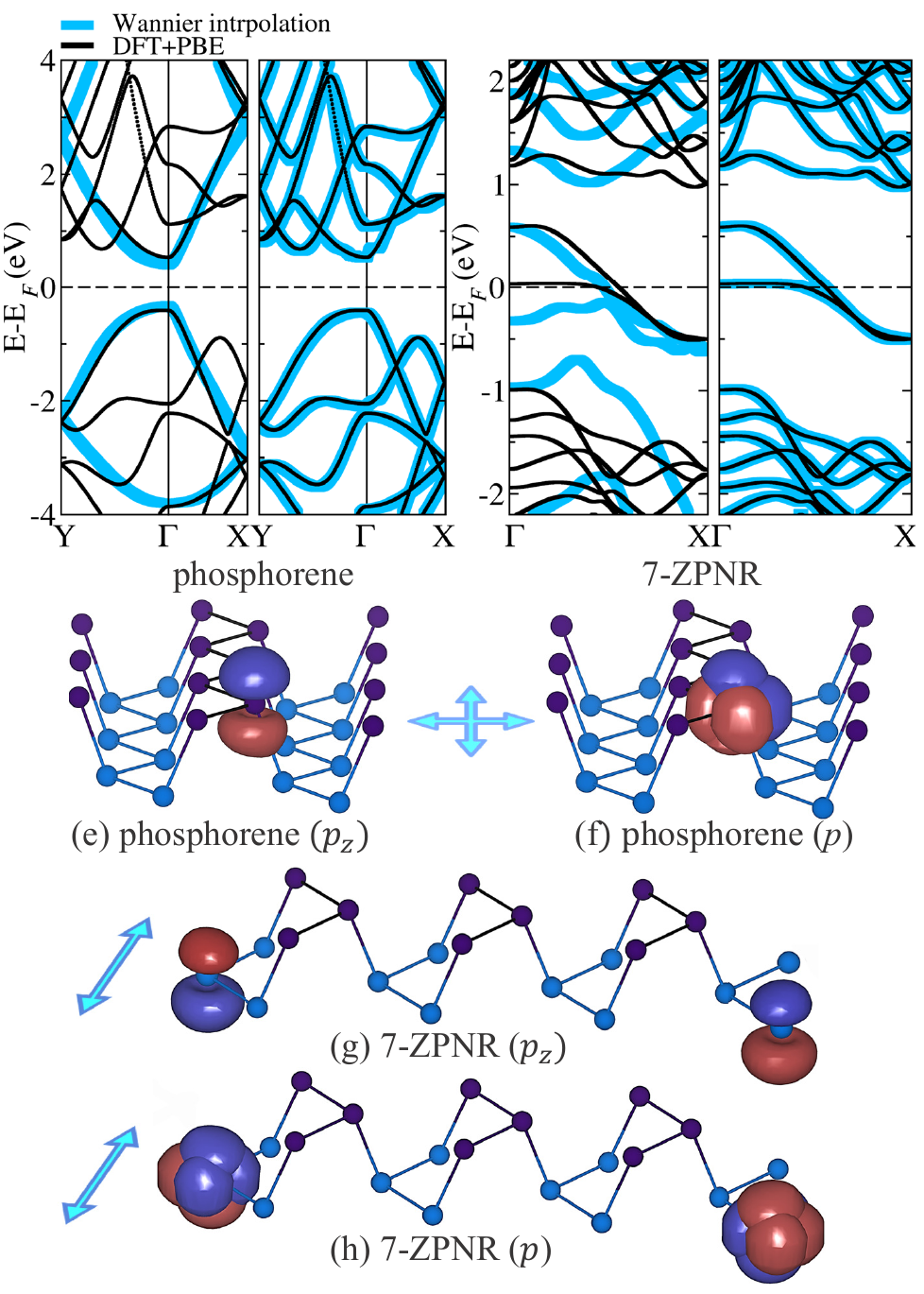}
	\vspace{-0.1 cm}
	\caption{(Color online) DFT-PBE and Wannier-interpolated band structure of
		 black phosphorene with (a) $p_z$ (b) $p$ correlated subspace, 7-ZPNR with (c) $p_z$ and (d) $p$ correlated subspace. (e)-(h) The corresponding $p_z$/$p$-like MLWF for P atoms of black phosphorene and 7-ZPNR.}
	\label{fig:subm3}
\end{figure}

where $L$ is the number of localized orbitals. The SPEX code uses the Wannier library to create the MLWFs for the $p$ orbitals of each atoms. For this construction, we used a dense $k$-point grid of $18\times18\times1$ ($8\times8\times8$) and $30\times1\times1$ for black phosphorene monolayer (bulk BP) and PNRs, respectively. Note that the matrix elements of the Coulomb potential are formally spin-dependent due to the spin dependence of the Wannier functions, we find that this dependence is negligible in practice. 
Note that polarization function largely depends
on the number of unoccupied states. We find that 50 bands per
atom in unit cell gives converged results. 
In calculation of polarization function we use Hilbert mesh with 500 frequencies (between 0 and 5 htr) and an accumulated stretching factor of 1.6 at 5 htr (see \cite{SPEX} for details). In the metallic systems to incorporate Drude term, we scales the head element of $W(k,\omega)$
 in the limit $k \rightarrow 0$ to enforce metallic screening.

\subsection{Correlated subspace}

To assess the magnitude of the screened Coulomb interaction, it is crucial to identify the correlated subspace. This subspace consists of electronic states that exhibit the strongest interactions, and its accurate determination is essential for constructing Wannier functions and the corresponding effective low-energy model Hamiltonian.
Although the dominant contribution to the bands near the Fermi level comes from the $p_z$ orbital, they are not well separated from the other bands with $p_x$ and $p_y$ character (see projected band structures in Figs.\,\ref{fig:subm2}). 
To validate the calculated Wannier functions and selected subspace, we compare the DFT-PBE and Wannier-interpolated band structures obtained using the set of $p$ and $p_{z}$ Wannier orbitals for black phosphorene and 7-ZPNR in Figs.\,\ref{fig:subm3}(a)-\ref{fig:subm3}(d). As observed, there is good overall agreement between the original and Wannier-interpolated bands in the $p$ correlated subspace, which is more suitable than the $p_z$ correlated subspace. The corresponding Wannier orbitals are depicted in Figs.\,\ref{fig:subm3}(e)-\ref{fig:subm3}(h) to indicate the well localized Wannier function. Thus, in this context, the full $p$ orbital set is the optimal correlated subspace for capturing the electronic characteristics of these structures.
However, for consistency with other results, we will also present the Coulomb interaction parameters for a one-orbital $p_z$ correlated subspace in some materials.

\section{Results and discussion}\label{sec3}

\subsection{black phosphorene and bulk black phosphorus}

We start by discussing the on-site and off-site effective Coulomb interaction parameters for black phosphorene and bulk BP. 
The results of bare, partially (cRPA), and fully (RPA) screened Coulomb interactions are presented in Table\,\ref{table1}. For comparison, we also report the corresponding values for pristine graphene. 
In the projected band structure of black phosphorene in Fig.\,\ref{fig:subm2}(a), it can be observed that the bands near the E$_F$ have predominantly $p_z$ orbital character, but they are not well separated from the other bands. Therefore, in addition to the $p_z$ orbitals, one can construct Wannier functions for the full $p$ shell orbitals as a correlated subspace. This situation is different in other planar 2D materials like graphene and h-BN, which do not exhibit buckling. In graphene, the $p_z$ band is largely decoupled from the other bands.

It is important to note that the choice of correlated subspace significantly affects the accuracy of specific properties under investigation~\cite{Karbalaee}.
For example, when calculating transport properties in phosphorene, a low-energy model considering only the single $p_z$ orbital may be sufficient, but to accurately investigate the  optical properties such as absorption and emission spectra, it is necessary to include the full $p$-orbital correlated subspace. Therefore, we will consider both a one-orbital ($p_z$) and a three-orbital ($p$) correlated subspace.
Furthermore, comparing these two subspaces will provide insight into the contribution of $p_z$ and $p$ electrons to the overall screening process.

The bare Coulomb interaction parameter $V$ provides information on the localization of Wannier functions.
Regardless of the correlated subspace, our calculated bare interaction for black phosphorene is almost 3.5 eV smaller than the bare Coulomb interactions of graphene~\cite{Ersoy}. This is expected, as the bare interaction generally decreases when moving downward in the periodic table from 2$p$ (graphene) to 3$p$ (phosphorene) materials, due to the lower degree of contraction of the 3$p$ wave functions compared to the 2$p$ ones.
As shown in Table\,\ref{table1}, when considering the $p$ subspace, the calculated on-site screened Coulomb interaction for phosphorene, denoted as $U_{00}$ (Hubbard $U$), is found to be 5.51 eV. However, when we consider the $p_z$ states as a correlated subspace,  the $p_x \rightarrow p_z$ and $p_y \rightarrow p_z$ channels contribute to the electronic polarization function. This leads to an enhancement of the dielectric screening and, consequently, a relatively smaller on-site $U$ value of 4.09 eV. Notably, this value is significantly smaller than the corresponding $U_{00}$ value in graphene, which is 8.7 eV~\cite{Wehling,Hadipour-1,Ersoy}.
When comparing graphene with phosphorene, each system contributes to the determination of the $U$ and $W$ parameters through two mechanisms.
Firstly, similar to the effect observed in the bare interaction, the spreading of the Wannier function leads to a reduction in $U$ and $W$ from graphene to phosphorene.
Secondly, the buckeling in black phosphorene brings the $p_x$ and $p_y$ states closer in energy proximity to the $p_z$ state, as shown in Fig.\,\ref{fig:subm2}(a). This proximity in energy levels increases the contribution of $p_x$/$p_y \rightarrow p_z$ transitions to the polarization function, resulting in a reduction of the $U$ and $W$ parameters.
In bulk phosphorene, the dielectric screening of Coulomb interaction is more efficient and occurs in all directions. As a result, the on-site Coulomb interaction obtained in bulk BP  is 4.99 eV, which is 0.5 eV smaller than the $U_{00}$ value of a single-layer phosphorene.
The difference between the partially screened $U$ (cRPA) and the fully screened interactions $W$ (RPA) determines the strength of the screening within the correlated subspace.
For semiconductors like phosphorene with the $p_z$ subspace, the difference $U-W$ is very small, turning out to be about 0.3 eV, which is approximately 7 percent of the $U$ value. However, for graphene the $U-W$ values are larger, ranging from 1.5 to 4.0 eV, which is about one-half of the $U$ value.
This indicates that $p_z \rightarrow p_z$ transitions are highly efficient for semimetallic systems like graphene and contribute significantly to the screening of the fully screened Coulomb interaction. On the other hand, due to the presence of a band gap in phosphorene, this screening effect diminishes, resulting in a very small $U-W$ difference.

\begin{figure}[b]
	\centering
	\includegraphics[width=75mm]{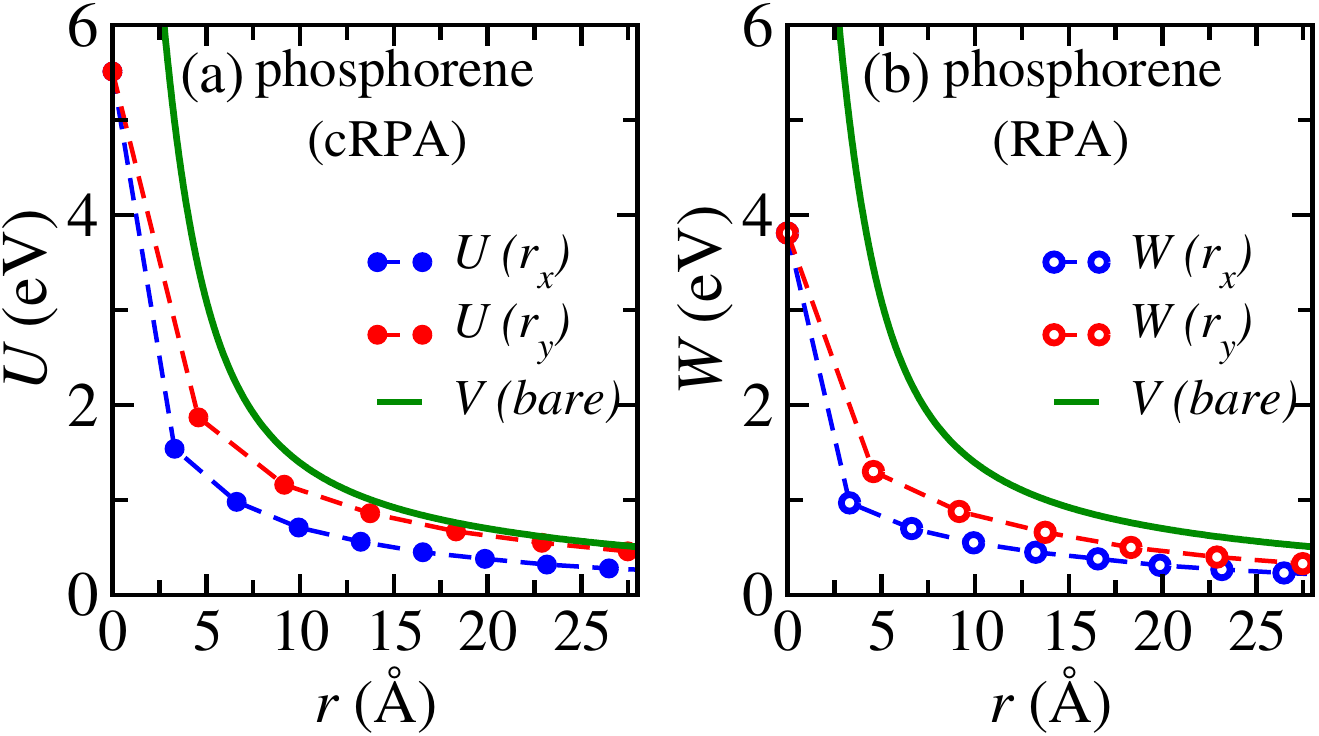}
	\vspace{-0.2 cm}
	\caption{(Color online) (a) partially $U$ (cRPA) (b) fully $W$ (RPA) screened inter-site Coulomb interaction for $p$ electrons as a function of distance $r$ along $x$ (zigzag) and $y$ (armchair) directions of black phosphorene. For comparison, the bare interaction $V(r)$ is also presented.}
	\label{fig:subm4}
\end{figure}

As shown in Table\,\ref{table1}, the semiconducting nature of phosphorene results in a significant long-range part of the effective Coulomb interaction, denoted as $U_{0,j}$. Furthermore, in Fig.\,\ref{fig:subm4}, the inter-site screened Coulomb interaction is indicated over $x$ and $y$ direction for cRPA and RPA approximation, as a function of distance $r$ between two P atoms. Overall, the screening is anisotropic and the Coulomb interactions in the zigzag ($x$) direction are less than those in the armchair ($y$) direction. We find sizable off-site $U$ and $W$ parameters at short distances, while at distances longer than 15 \AA, the interaction is unscreened in the armchair direction [see Fig.\,\ref{fig:subm4} (a)]. These findings are consistent with substantial $GW$ corrections, which yield a quasi-particle band gap of 2.2 eV, as well as experimental observations of highly anisotropic excitons with a considerable binding energy of approximately 0.9 eV in black phosphorene~\cite{Wang,Maciej}.
Graphene has no band gap, and it is expected that its long-range interaction is less than in phosphorene, which is a semiconductor. However, the results show the opposite. This peculiar screening can be attributed to the zero-gap chiral relativistic nature of graphene~\cite{Wehling,Hwang}. From the point of view of screening, electrons in linear bands behave similarly to those in an insulator, and therefore do not screen well long-range Coulomb interactions.  In bulk BP, the situation is different due to the increased coordination number. As a result, the off-site effective Coulomb interaction decreases more rapidly compared to phosphorene.

\begin{figure}[t]
	\centering
	\includegraphics[width=78mm]{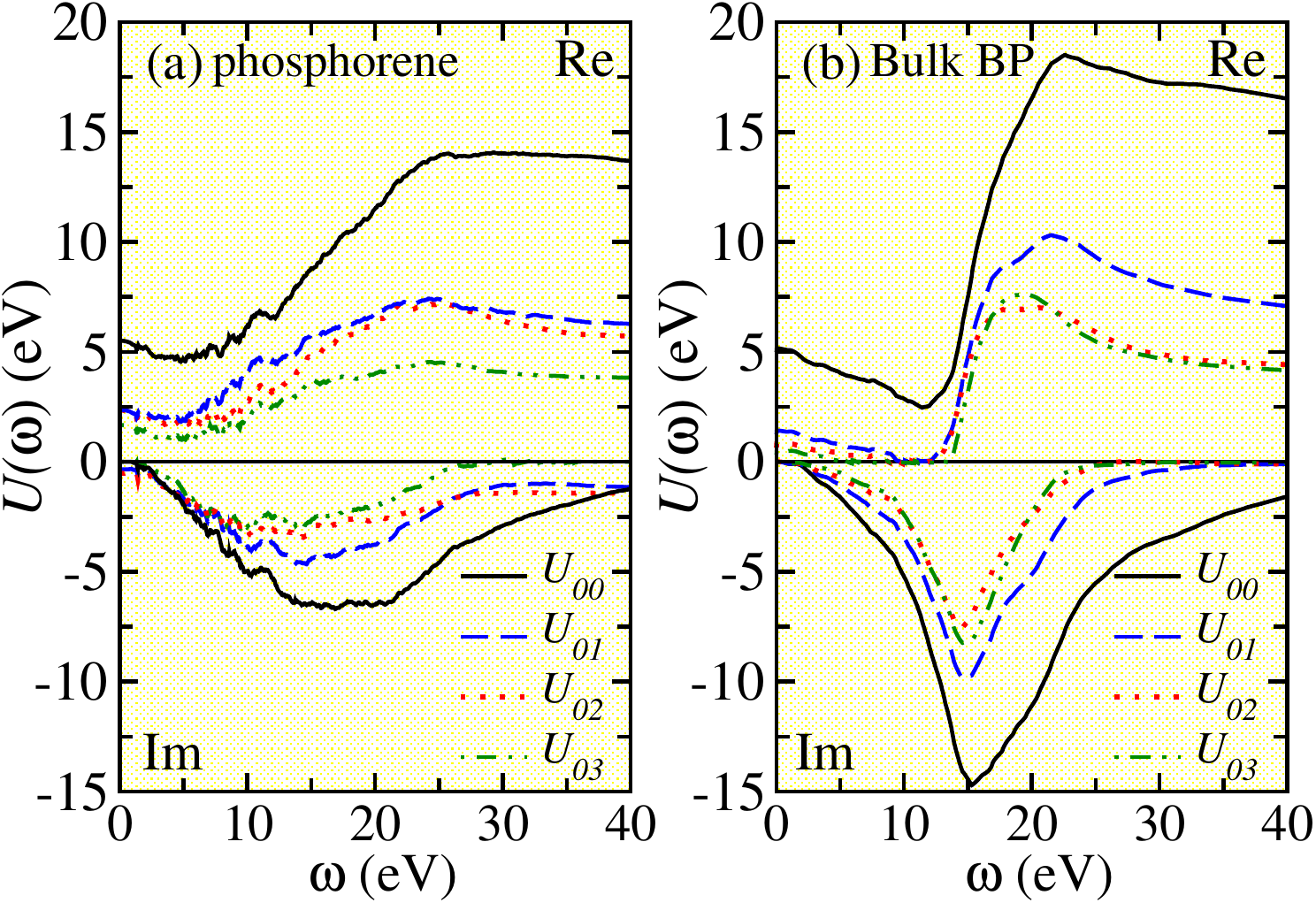}
	\vspace{-0.1 cm}
	\caption{(Color online) Frequency dependence of the onsite and off-site Coulomb interaction parameters $U(\omega)$ for (a)
black phosphorene and (b) Bulk BP. The real and imaginary parts
of $U(\omega)$ are presented individually.}
	\label{fig:submx}
\end{figure}

At the end of this section, we examine the frequency-dependent behavior of the partially $U(\omega)$. Our focus is on black phosphorene and bulk black phosphorus, analyzing these materials with respect to $p$ as the correlated subspaces. The real and imaginary components of the computed on-site interaction $U_{00}(\omega)$, along with the interactions of the first, second, and third nearest neighbors (namely $U_{01}(\omega)$, $U_{02}(\omega)$, and $U_{03}(\omega)$), are depicted in Fig.~\ref{fig:submx} for these materials.
In the case of black phosphorene, the $U(\omega)$ profile demonstrates smooth behavior with minor fluctuations up to a frequency of 10 eV. Beyond this threshold, it shows linear growth, reaching a peak at the plasmon frequency of around 20 eV (also noticeable as a valley in the imaginary part of $U(\omega)$). Subsequently, as the frequency continues to rise, it gradually approaches the static value of 13.2 eV specific to phosphorene. The scenario remains similar for the off-site (nearest-neighbor $U_{0j}(\omega)$) Coulomb interaction, indicated by dashed lines in Fig.~\ref{fig:submx}(a). It is important to highlight that due to the minimal variation of the effective Coulomb interaction for 2D phosphorene at low frequencies, utilizing the static $U(\omega=0)$ in model Hamiltonian studies could be reliable.

For the bulk black phosphorus system, the Coulomb interactions exhibit a decrease in $U(\omega)$ at low frequencies between 0-10 eV, stemming from effective screening influenced by more $p$ states around the Fermi level. Notably, the frequency dependence of $U_{00}(\omega)$ mirrors similar trends seen in the $U_{01}(\omega)$, $U_{02}(\omega)$, and $U_{03}(\omega)$ cases.  Hence, relying on the static limit $U(\omega=0)$ for model Hamiltonian studies of bulk systems may not be appropriate.
Lastly, it's worth mentioning that the imaginary part of $U(\omega)$ offers qualitative insights into the plasmon frequency and damping. As anticipated, the peaks of the imaginary part $U(\omega)$ are broader in 2D black phosphorene compared to bulk black phosphorus, indicating more pronounced plasmon damping.

\subsection{Hydrogen-passivated PNRs: armchair and zigzag edges}

\begin{figure}[t]
	\centering
	\includegraphics[width=86mm]{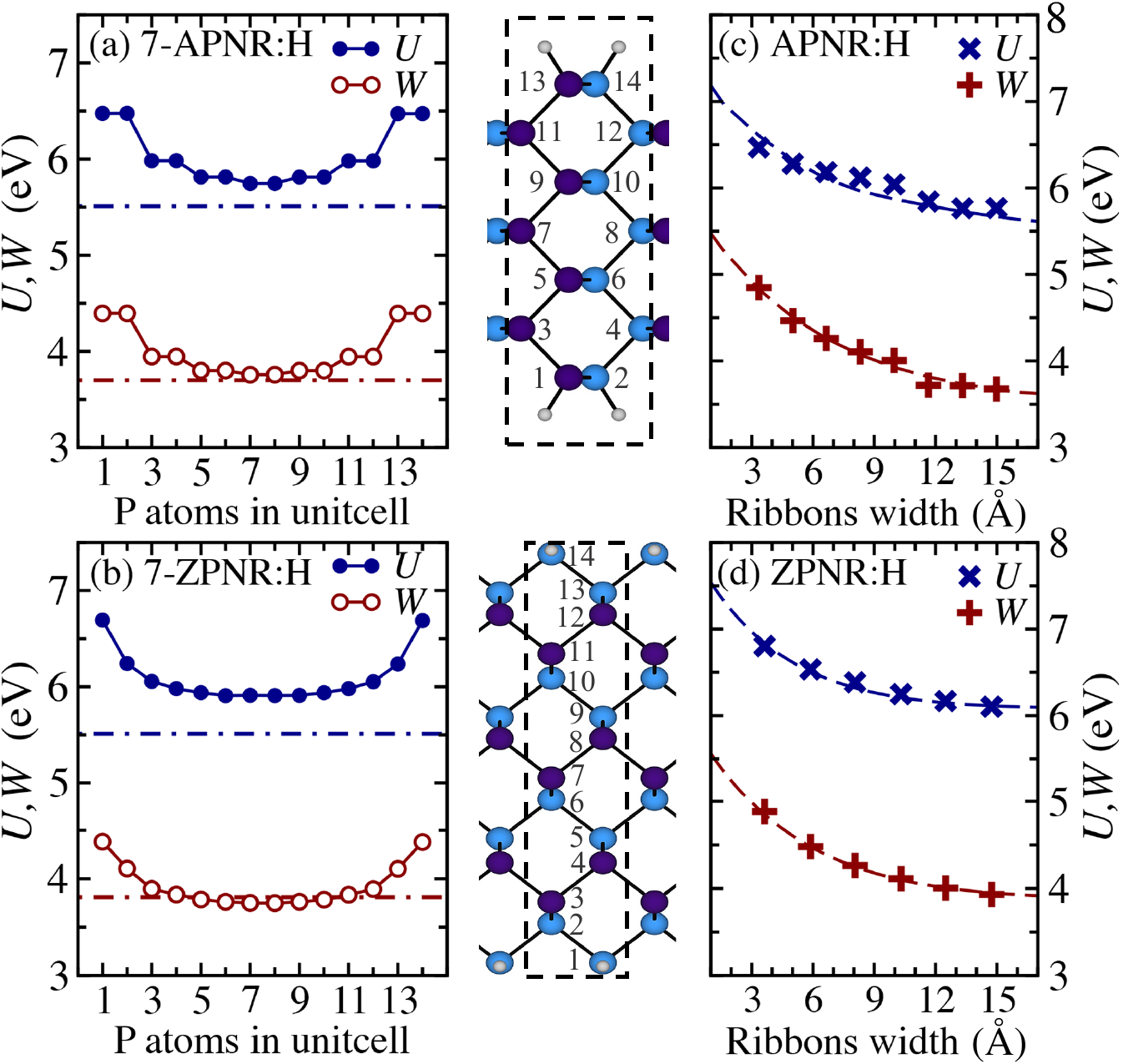}
	\vspace{-0.6 cm}
	\caption{(Color online) Calculated partially screened on-site Coulomb interaction $U$ (blue line) and fully screened Coulomb interaction $W$ (red line) of $p$ electrons  for (a) 7-APNR:H, (b) 7-ZPNR:H (the index of P atoms in unit cell is shown in Middle panel). The average effective Coulomb interaction parameters in both (c) armchair and (d) zigzag H-passivated PNRs. For comparison, the $U$ and $W$ values of pristine black phosphorene are also presented (dashed lines).}
	\label{fig:subm5}
\end{figure}

In this section, we investigate the matrix elements of the effective Coulomb interaction for hydrogen-passivated PNRs:H with both armchair and zigzag edges. The results of DFT-PBE calculations indicate that both the APNRs:H and ZPNRs:H exhibit semiconducting behavior. Figs.\,\ref{fig:subm5}(a) and \,\ref{fig:subm5}(b) show the variation of the on-site Coulomb interaction for the phosphorus atoms across the ribbons of 7-APNR:H and 7-ZPNR:H, respectively. For comparison, the corresponding $U$ and $W$ values for pristine black phosphorene with  correlated subspace $p$ are presented using dashed lines. Generally, when the dimensionality of materials is reduced and the quantum confinement effect is intensified, the effective Coulomb interaction parameters are expected to be enhanced. Taking the example of 7-APNR:H, the calculated values of $U$ ($W$) turn out to be around 6.04 eV (4.01 eV), compared to 5.51 eV (3.81 eV) for pristine black phosphorene and 4.99 eV (3.73 eV) for bulk BP. Additionally, notable $GW$ corrections to the band gap have been obtained in PNRs~\cite{Nourbakhsh}, with the values ranging from 2.1 eV to 4.4 eV in ZPNR:H and from 1.2 eV to 3.0 eV in APNR:H with 1.0 nm width, which is consistent with their significant effective Coulomb matrix elements.

\begin{table*}[t]
	\caption{The value of parameters related to antiscreening in PNRs, such as ribbons width, $r_{c_{1}}$ (transition from screening to antiscreening) and $r_{c_{2}}$ (transition from antiscreening to unscreening), antiscreening zone of $\delta r_{c}$=$r_{c_{2}}$-$r_{c_{1}}$ in {\AA}, and the antiscreening contribution of $\Delta$ (the maximum amplitude of the difference $V - W$ within antiscreening range) for edge and central atoms of $N_{a}$=3 to 7 in $N_{a}$-APNR:H and	$N_{z}$=2 to 5 in $N_{z}$-ZPNR:H. Furthermore, corresponding values for AGNR:H and \emph{h}-BNNR:H are presented for comparison.}
	\centering
	\begin{ruledtabular}
		\begin{tabular}{ccccccccccccc}
			&  &   &  & \multicolumn{4}{c}{Inner} &  & \multicolumn{4}{c}{Edge}  \\ \cline{5-8} \cline{10-13} 
			\rule{0pt}{5mm}%
			Systems&&Ribbon Width (\AA)&&$r_{c_{1}}$ (\AA)&$r_{c_{2}}$ (\AA)&$\delta r_{c}$ (\AA)& $\Delta$ (eV)&&$r_{c_{1}}$ (\AA)&$r_{c_{2}}$ (\AA)&$\delta r_{c}$ (\AA)&$\Delta$ (eV)\\ 
			\hline
			\rule{0pt}{5mm}%
			3-APNR:H   &&3.4&&16.9&85.3&68.4&-0.024&&18.0&84.3&66.3&-0.022\\
			4-APNR:H   &&5.0&&23.6&78.8&55.2&-0.017&&26.3&75.7&49.4&-0.013\\
			\textbf{5}-APNR:H&&\textbf{6.6}&&28.5&72.3&\textbf{43.8}&\textbf{-0.012}&&36.1&62.3&\textbf{26.2}&\textbf{-0.004}\\
			6-APNR:H   &&8.3&&36.4&64.8&28.4&-0.005&& -  & -  & -  &  -  \\
			7-APNR:H   &&10.0&&48.7&48.7&00.0&$\approx$0.000&& -  & -  & -  &  -  \\
			\rule{0pt}{4mm}%
			2-ZPNR:H   &&3.7&&12.8&69.4&56.6&-0.055&&13.0&69.0&56.0&-0.053\\
			\textbf{3}-ZPNR:H&&\textbf{6.3}&&18.1&60.4&\textbf{42.3}&\textbf{-0.029}&&19.5&58.5&\textbf{39.0}&\textbf{-0.023}\\
			4-ZPNR:H   &&8.5&&24.6&51.6&27.0&-0.011&&29.9&44.9&15.0&-0.003\\
			5-ZPNR:H   &&10.8&&39.4&39.4&00.0&$\approx$0.000&&-&-&- & -\\
			\rule{0pt}{4mm}%
			3-APNR     &&3.8&&26.3&94.0&67.7&-0.057&&26.6&93.6&67.0&-0.055\\
			4-APNR     &&5.5&&21.5&90.0&42.3&-0.042&&21.8&89.4&67.6&-0.040\\
			5-APNR     &&7.2&&24.7&88.4&63.7&-0.039&&25.1&87.9&62.8&-0.037\\
			6-APNR     &&8.9&&33.3&71.8&36.5&-0.011&&34.7&66.8&32.1&-0.009\\
			7-APNR     &&10.6&&53.2&53.2&00.0& 0.000&&53.2&53.2&00.0&$\approx$0.000\\ \hline
			\rule{0pt}{5mm}%
			\textbf{6}-AGNR:H\cite{Hadipour}&&\textbf{6.2}&&22.0&115.0&\textbf{93.0}&\textbf{-0.047}&&25.1&114.5&\textbf{89.4}&\textbf{-0.037}\\
			7-AGNR:H\cite{Hadipour}&&7.3&&22.0&110.0&88.0&-0.036&&25.6&109.0&83.4&-0.026\\
			8-AGNR:H\cite{Hadipour}&&8.6&&35.0&105.0&70.0&-0.032&&42.6&102.3&59.7&-0.015\\
			9-AGNR:H\cite{Hadipour}&&9.9&&23.0&65.0 &42.0&-0.041&&28.0&60.0 &32.0&-0.016\\ 
			10-AGNR:H&&11.1&&15.0&23.0&8.0&-0.02&& -& - & - &  -\\ \hline
			\rule{0pt}{5mm}%
			3-Ah-BNNR:H\cite{Montaghemi}&&2.4&& 8.3&90.0 &81.7&-0.040&&9.7 &90.0 &80.3&-0.028 \\
			4-Ah-BNNR:H\cite{Montaghemi}&&3.7&&10.0&88.0 &78.0&-0.025&&11.2&84.0 &72.8&-0.016\\
			5-Ah-BNNR:H\cite{Montaghemi}&&5.0&&12.5&86.0 &73.5&-0.018&&16.5&81.0 &64.5&-0.013\\
			\textbf{6}-Ah-BNNR:H\cite{Montaghemi}&&\textbf{6.2}&&17.0&85.0 &\textbf{68.0}&\textbf{-0.014}&&23.0&76.0 &\textbf{53.0}&\textbf{-0.008}\\
			\rule{0pt}{4mm}%
			3-Zh-BNNR:H\cite{Montaghemi}&&5.0&&10.5&49.0 &38.5&-0.037&&12.3&42.5 &30.2&-0.025\\
			4-Zh-BNNR:H\cite{Montaghemi}&&7.2&&13.5&45.0 &31.5&-0.022&&15.8&41.2 &25.4&-0.013\\
			5-Zh-BNNR:H\cite{Montaghemi}&&9.4&&16.5&41.0 &24.5&-0.012&&-   & -   & -  &  -
		\end{tabular} 
		\label{table2}
	\end{ruledtabular}
\end{table*}

\begin{figure}[b!]
	\centering
	\includegraphics[width=76mm]{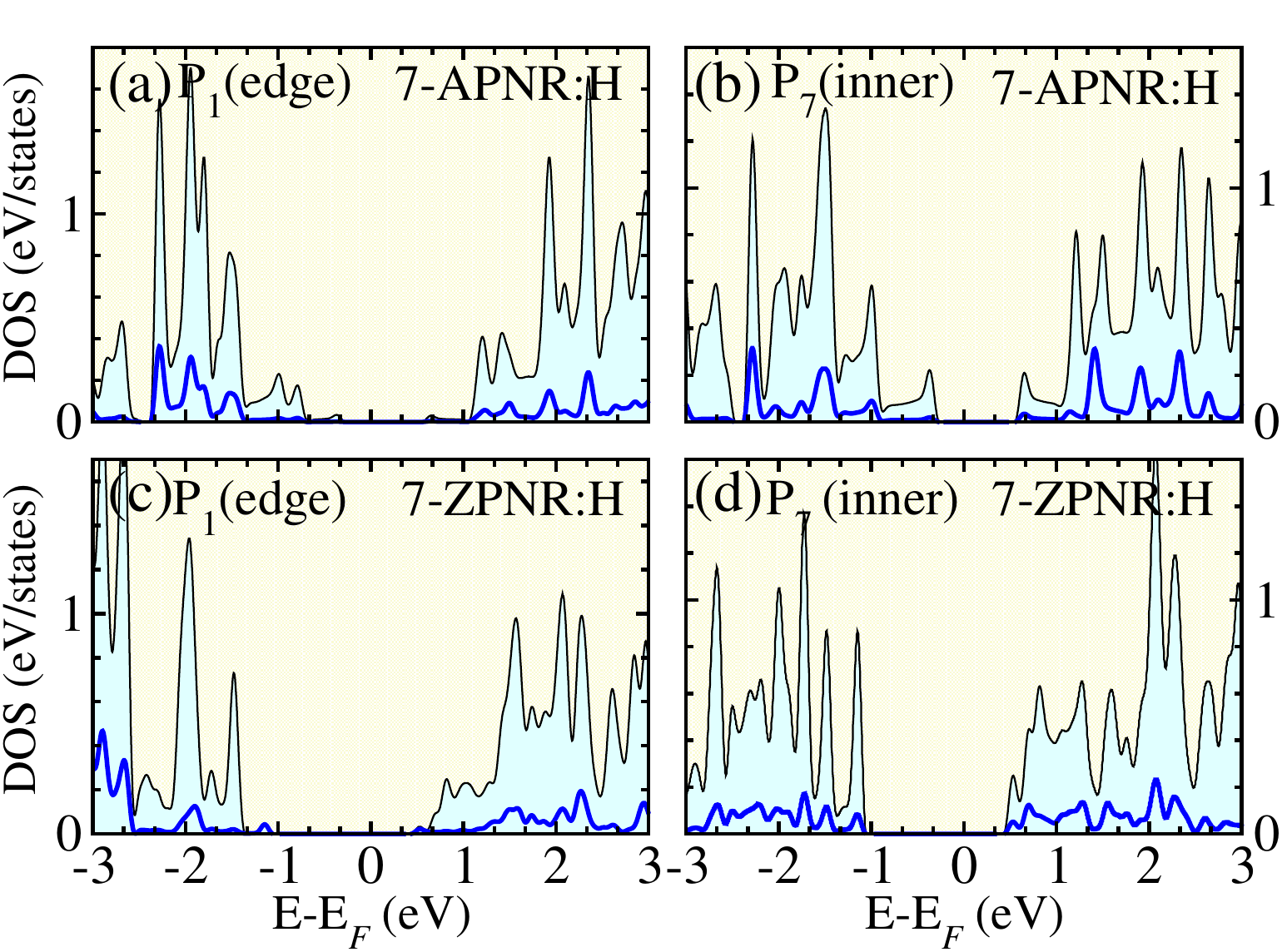}
	\vspace{-0.2 cm}
	\caption{(Color online) Projected density of states for (a) edge, (b) inner P atoms of 7-APNR:H and (c) edge, (d) inner P atoms of 7-ZPNR:H.}
	\label{fig:subm6}
\end{figure}

In both types of hydrogen-passivated PNR with armchair and zigzag edge, the $U$ and $W$ parameters for the edge atoms of the nanoribbons significantly increase compared to the corresponding values for the inner atoms. The situation is more and less the same for other PNRs:H. This result contrasts with semiconducting AGNRs, where the effective Coulomb interaction parameters across the ribbon are nearly identical~\cite{Hadipour}. To explain the sizable effective interaction of edge atoms, we obtained the projected density of states (PDOS) for the edge and inner P atoms of 7-APNR:H and 7-ZPNR:H. As shown in Figs.\,\ref{fig:subm6}(c)-\ref{fig:subm6}(f), the $p$ states move away from E$_F$ and create a larger band gap when transitioning from inner atoms to edge atoms. Consequently, the electronic screening, resulting from the contribution of $s \rightarrow p$ and $s \rightarrow s$ transitions, is considerably reduced at the edge atoms. It should be noted that this variation in the electronic structure is due to the large structural puckering in PNR, which is not present in AGNRs.

Previous experimental and theoretical studies on graphene~\cite{Hadipour,Magda,Son1}, black phosphorene~\cite{Watts,Wu,Han,Nourbakhsh}, and \emph{h}-BN nanoribbons~\cite{Montaghemi} revealed a close relationship between the nanoribbon width and its electronic properties. It has been predicted that the band gaps of hydrogen-passivated PNRs with armchair and zigzag edge decrease monotonically with increasing ribbon width~\cite{Wu}. To illustrate the effect of quantum confinement on the electronic screening of Coulomb interaction, we also consider the average $U$ and $W$ parameters of P atoms as a function of the ribbon width for armchair and zigzag hydrogen-passivated PNRs. We consider several PNRs:H with different widths, i.e., $N_{a}$=3 to 10 with armchair edges and $N_{z}$=2 to 8 with zigzag edges. As shown in Figs.\,\ref{fig:subm5}(c) and \,\ref{fig:subm5}(d), similar to the band gaps~\cite{Han}, the average effective Coulomb interaction parameters in both armchair and zigzag PNRs decrease with increasing ribbon widths, and we see the slightly larger screened parameters in zigzag PNRs. These results contrast with AGNRs, which show oscillatory behavior of the band gaps and average screened parameters~\cite{Hadipour}.

\begin{figure}[t]
	\centering
	\includegraphics[width=87mm]{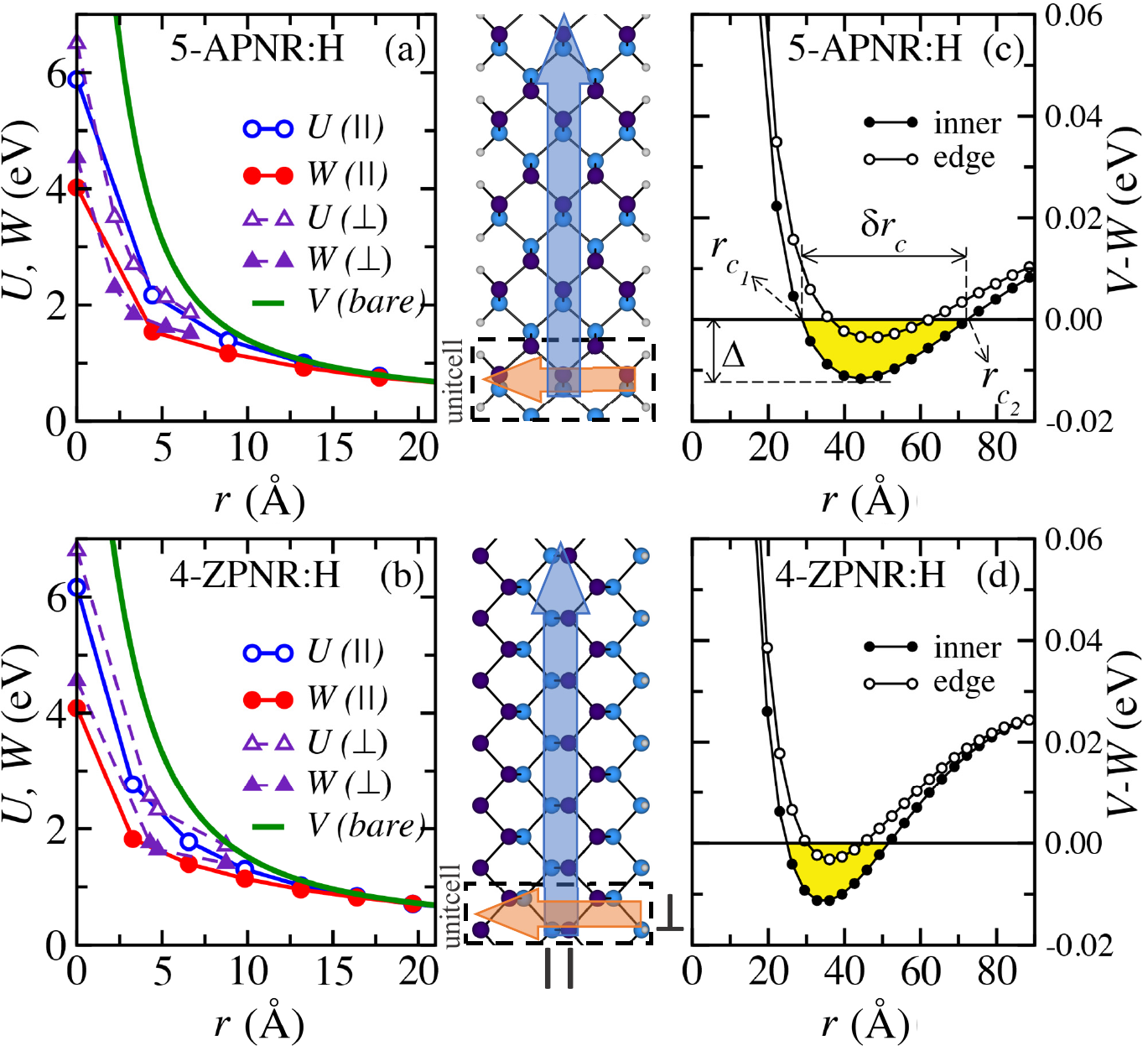}
	\vspace{-0.7 cm}
	\caption{(Color online) partially (fully) screened inter-site Coulomb interaction $U$ ($W$) for $p$ electrons as a function of distance $r$ for hydrogen-passivated (a) APNR (5-APNR:H) and (b) ZPNR (4-ZPNR:H). Here, the symbols $(\parallel)$ and $(\perp)$ correspond to interactions along the ribbon and across the ribbon, respectively. The bare interaction $V (r)$ is presented with the green line. The difference $V - W$ between bare and fully screened interaction as a function of distance $r$ between two P atoms along the ribbon for the edge and central atoms of (c) 5-APNR:H and (d) 4-ZPNR:H.}
	\label{fig:subm7}
\end{figure}

To investigate how quantum confinement affects the screening of long-range Coulomb interactions in black phosphorene, we present the off-site partially (fully) Coulomb interaction parameters, along the ribbon $U_{\parallel}$ ($W_{\parallel}$), and across the ribbon $U_{\perp}$ ($W_{\perp}$), as a function of distance $r$ between two P atoms for 5-APNR:H and 4-ZPNR:H  in Figs.\,\ref{fig:subm7}(a) and \,\ref{fig:subm7}(b). The green line represents the bare Coulomb interaction $V$ for comparison. Our results indicate that the Coulomb interaction is well screened at short distances, approximately two lattice spacing, and unscreened after that in both hydrogen-passivated PNRs with armchair and zigzag edges. The difference $V – W$ gradually decreases as the distance $r$ increases.
To examine the behavior of the Coulomb interaction at larger distances, we present the difference $V – W$ for the edge (white circle) and inner (black circle) atoms of 5-APNR:H and 4-ZPNR:H in Figs.\,\ref{fig:subm7}(c) and \,\ref{fig:subm7}(d), respectively, up to 90 {\AA}. When $r$ exceeds 25 {\AA}, $V – W$ becomes negative, indicating antiscreening. This phenomenon has been reported in previous theoretical studies on low-dimensional insulators such as AGNRs~\cite{Hadipour} and \emph{h}-BNNRs~\cite{Montaghemi}, zero-dimensional Nb$_4$Co clusters~\cite{Peters}, as well as carbon nanotubes~\cite{Deslippe}. As we continue calculating the $W$ parameters over longer distances, $V – W$ returns to the positive region, and the Coulomb interaction gradually becomes unscreened. 
For example, in the case of central atoms in 5-APNR:H, we observe antiscreening between the critical points $r_{c_{1}}$ = 28.5 {\AA} (transition from screening to antiscreening) and $r_{c_{2}}$ = 72.3 {\AA} (transition from antiscreening to unscreening). Similarly, for 4-ZPNR:H, the corresponding values are 24.6 {\AA} and 51.6 {\AA}, respectively. Also, results indicate that antiscreening slightly decreases when moving from inner to edge atoms in both cases.

The concept of antiscreening was validated in one-dimensional semiconductors and large molecules~\cite{Deslippe,Brink}. Put simply, when an electron is influenced by the electric field of another electron, the surrounding medium adjusts the other charges to diminish the bare interaction between the two electrons, a process known as screening. Conversely, antiscreening arises when the medium amplifies the direct interaction between the two electrons.
To comprehend how antiscreening occurs, one can envision the medium as comprising point dipoles surrounding two point charges. These dipoles can be categorized into screening dipoles and antiscreening dipoles. Dipoles positioned between the charges contribute to the amplification, or antiscreening, of the bare interaction, while the other surrounding dipoles diminish, or screen, this interaction. In the context of one-dimensional systems, the proportion of the antiscreening area (space between the charges) to the external area is crucial. This is why antiscreening is observed in one-dimensional systems like carbon nanotubes, large organic molecules, and clusters~\cite{Peters,Rossen}.
Note that $V-U$ eventually returns to positive region at a large distance $r_{c_{2}}$. This is because the antiscreening in nanoribbons takes place in quasi-one dimension rather than precisely one dimension. In an ideal one-dimensional system, $r_{c_{2}}$ approaches infinity, and there is no return to the positive region at a large distance. That is why $r_{c_{2}}$ increases as the ribbon width decreases [see Table\,\ref{table2}].

To investigate the effect of the nanoribbon’s width on antiscreening, we extend our calculations to wider nanoribbons. Table\,\ref{table2} presents additional values related to antiscreening in PNRs, including the ribbon’s width, critical points $r_{c}$, the antiscreening zone $\delta r_{c}$ ($r_{c_{2}}$ - $r_{c_{1}}$), and the maximum value of antiscreening $\Delta$ for both edge and inner atoms. The Table\,\ref{table2} includes data for nanoribbons with $N_{a}$ ranging from 3 to 7 and $N_{z}$ ranging from 2 to 5. For comparison, we also provide corresponding values for AGNRs and \emph{h}-BNRs~\cite{Hadipour,Montaghemi}.
As expected, both APNRs:H and ZPNRs:H exhibits a decrease in the values of $\Delta$ and $\delta r_{c}$ as the ribbon width increases. For widths greater than 10.8 {\AA}, no antiscreening is observed. This threshold is similar to the corresponding values for \emph{h}-BNNRs and AGNRs, which are 12.6 {\AA} and 11.1 {\AA}, respectively ~\cite{Hadipour,Montaghemi}. This suggests that the antiscreening effect starts when the ribbon width is less than approximately 1 nm in quasi-one-dimensional nanoribbons. 

We find anisotropic behaviors between the APNRs and ZPNRs. If we consider two 5-APNR:H and 3-ZPNR:H with the same width for comparison. As seen in Table\,\ref{table2}, the values of $\Delta$ and $\delta r_{c}$ in the zigzag type are higher than those of armchair ones, especially at the edges of nanoribbons. So, stronger antiscreening and the larger on-site screened parameters in the zigzag type of PNRs are in agreement with a larger exciton binding energy and stronger quantum confinement of ZPNR:H in the same width as APNR:H, according to previous \emph{ab initio} calculations~\cite{Nourbakhsh}. 
Furthermore, the antiscreening contribution $\Delta$ in PNRs and \emph{h}-BNNRs is nearly identical but less than that in GNRs. This decrease in antiscreening strength in APNRs can be attributed to the puckering of the black phosphorene lattice, in contrast to the hexagonal planar structure of graphene.

We find that the inter-site Coulomb interaction at longer distances is barely screened or even antiscreened in PNRs:H and a sizable on-site effective interaction at the nanoribbon's edge. This reduced screening of the local and non-local Coulomb interactions can be attributed to the strong quantum confinement and enhancement in the overlap between electron and hole wave functions as the ribbon width decreases. One consequence of this reduced screening is the formation of tightly bound excitons, which has been experimentally observed in PNRs~\cite{Biswas,Watts,Macdonald,Liu2}. photoluminescence excitation spectroscopy suggests an edge-confined exciton with the instrument's limited lifetime of $\sim$100 ps and binding energy of $\sim$1.7 eV~\cite{Biswas}. As well as theoretical studies have reported corresponding binding energies of approximately 1.6 eV and 1.4 eV for 1 nm width in zigzag and armchair PNRs, respectively~\cite{Nourbakhsh}. This decrease in dielectric screening of the Coulomb interaction, caused by the reduced dimensionality and strong excitonic states, has also been observed in other low-dimensional systems such as graphene~\cite{Cudazzo}, AGNRs~\cite{Denk,Prezzi1,Prezzi2,Soavi}, fluorographene~\cite{Karlicky}, and transition-metal dichalcogenides~\cite{He,Ugeda,ebe_3,ebe_4,ebe_5}. Also, the experimental extraction of the excitonic ladder in phosphorene reveals significant adjustments to the energy ladder of excitonic states, demonstrating a notable deviation from the conventional hydrogenic Rydberg series~\cite{Maciej}. Therefore, the existence of tightly bound excitons, and the nonhydrogenic behavior of the excitonic states, revealing a significantly reduced and nonlocal dielectric screening of the Coulomb interaction in these materials.

\begin{figure}[b]
	\centering
	\includegraphics[width=87mm]{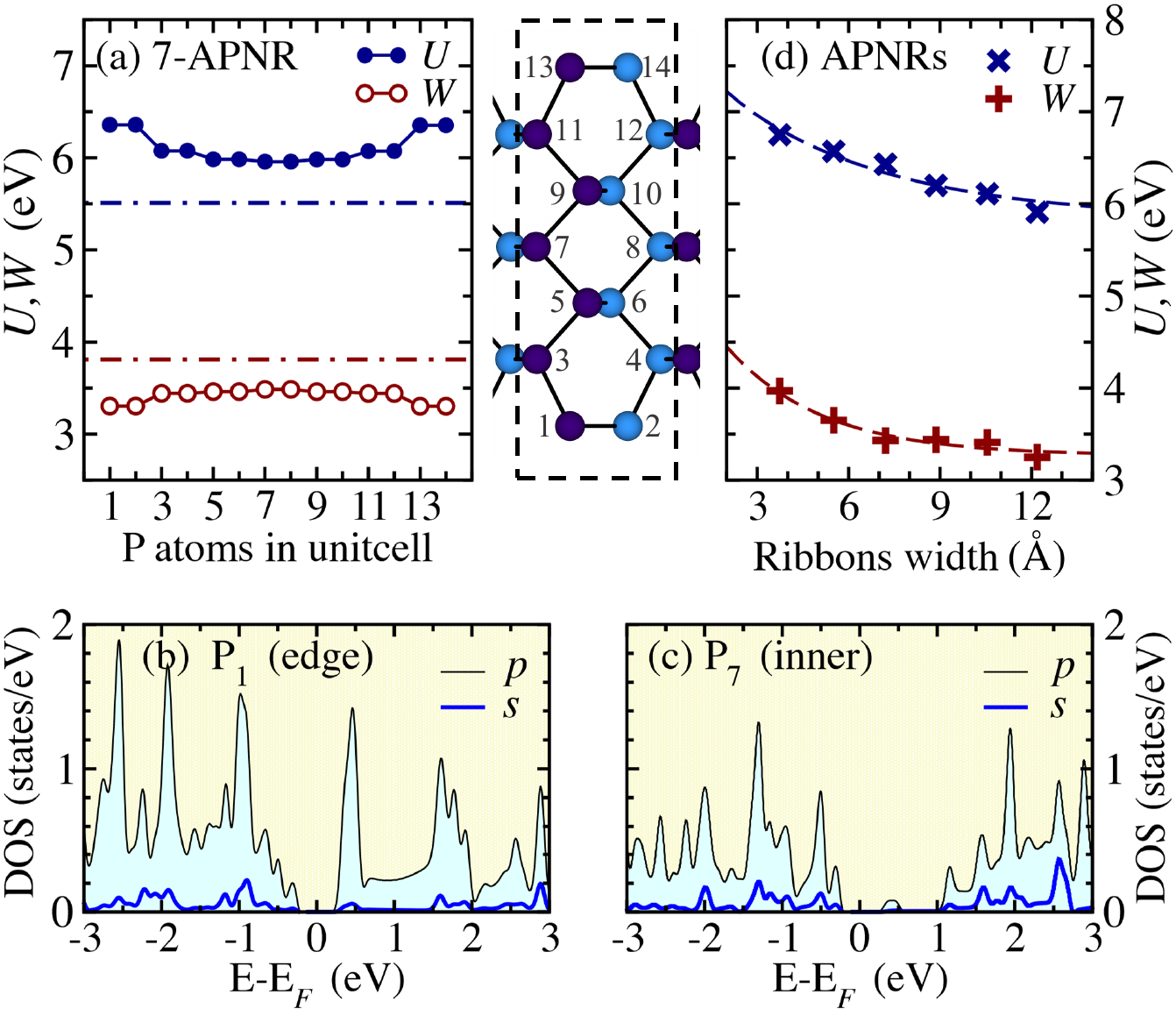}
	\vspace{-0.6 cm}
	\caption{(Color online) (a) On-site Coulomb interaction $U$ (blue line) and fully screened Coulomb interaction $W$ (red line) of $p$ electrons  for a bare APNRs (7-APNR). Projected density of states for (b) edge and (c) inner P atoms of 7-APNR. (d) The average effective Coulomb interaction parameters versus ribbon widths for bare APNRs.}
	\label{fig:subm8}
\end{figure}
    \subsection {Bare armchair phosphorene nanoribbons}

To investigate the effect of hydrogenation of ribbon edges on the short and long-range Coulomb interaction parameters, we will focus on unsaturated edge PNRs. Bare APNRs are semiconductors, while ZPNRs are metallic systems, which will be discussed in detail in the next section.
As an example, Fig.\,\ref{fig:subm8}(a) illustrates the variation of the on-site Coulomb interaction parameters for phosphorus atoms across the ribbon width in 7-APNR. Hubbard $U$ interactions of bare APNRs are more or less the same as corresponding ones for edge hydrogenated APNRs:H. However, the fully screened parameters $W$ are significantly reduced. This behavior can be explained by examining the PDOS shown in Figs.\,\ref{fig:subm8}(b) and \,\ref{fig:subm8}(c) for the edge and inner P atoms, respectively.
The increase in the $p$ states of P atoms near E$_{F}$ results in their significant contribution to the total polarization function. Consequently, electronic screening is enhanced due to the $p$ $\rightarrow$ $p$ transitions, leading to a reduction in the $W$ parameter. This effect is more pronounced at the edge, resulting in smaller $W$ values.

\begin{figure}[t]
	\centering
	\includegraphics[width=87mm]{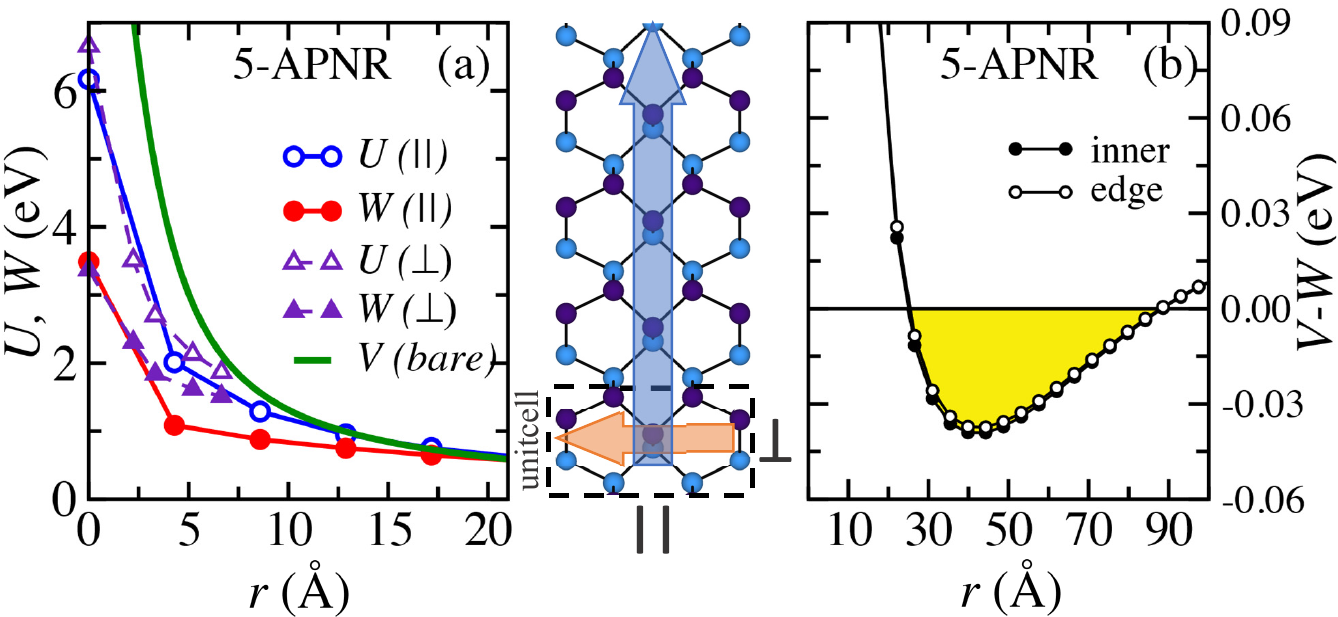}
	\vspace{-0.5 cm}
	\caption{(Color online)  (a) partially (fully) screened inter-site Coulomb interaction $U$ ($W$) for $p$ electrons as a function of distance $r$ between two P atoms along $(\parallel)$ and acrros $(\perp)$ the ribbon for bare APNRs (5-APNR). The bare interaction $V (r)$ is presented with the green line. (b) The difference $V - W$ for edge and central atoms of 5-APNR.}
	\label{fig:subm9}
\end{figure}

Furthermore, in Fig.\,\ref{fig:subm8}(d), we present the average $U$ and $W$ parameters of the P atoms as a function of the ribbon width for bare armchair PNRs. Results related to bare armchair PNRs are quite similar to those in hydrogen-passivated PNRs, where the average effective Coulomb interaction parameters decrease monotonically with increasing ribbon width.
In Fig.\,\ref{fig:subm9}(a), the same trend can be observed for the long-range effective Coulomb interaction, with the difference that the value of $W$ for bare APNRs is somewhat smaller than that of the hydrogen-passivated system.
According to Fig.\,\ref{fig:subm9}(b), antiscreening is observed between the critical points $r_{c_{1}}$=25 {\AA} and $r_{c_{2}}$=88 {\AA} which does not change by moving from inner to edge atoms. A comparison between Figs.\,\ref{fig:subm9}(b) and \,\ref{fig:subm7}(c) reveals that the antiscreening zone $\delta r_{c}$ and the antiscreening contribution $\Delta$ in 5-APNR are significantly larger than those in 5-APNR:H. Therefore, hydrogen passivation reduces the extent of antiscreening in PNRs.
Furthermore, Table\,\ref{table2} displays the values of $\delta r_{c}$ and $\Delta$ for $N_{a}$=3 to 7 in APNRs. No antiscreening occurs for widths greater than 10.6 {\AA}, which is slightly larger than the corresponding hydrogen-passivated APNRs.

\subsection{Bare zigzag phosphorene nanoribbons}

\begin{figure}[t]
	\centering
	\includegraphics[width=87mm]{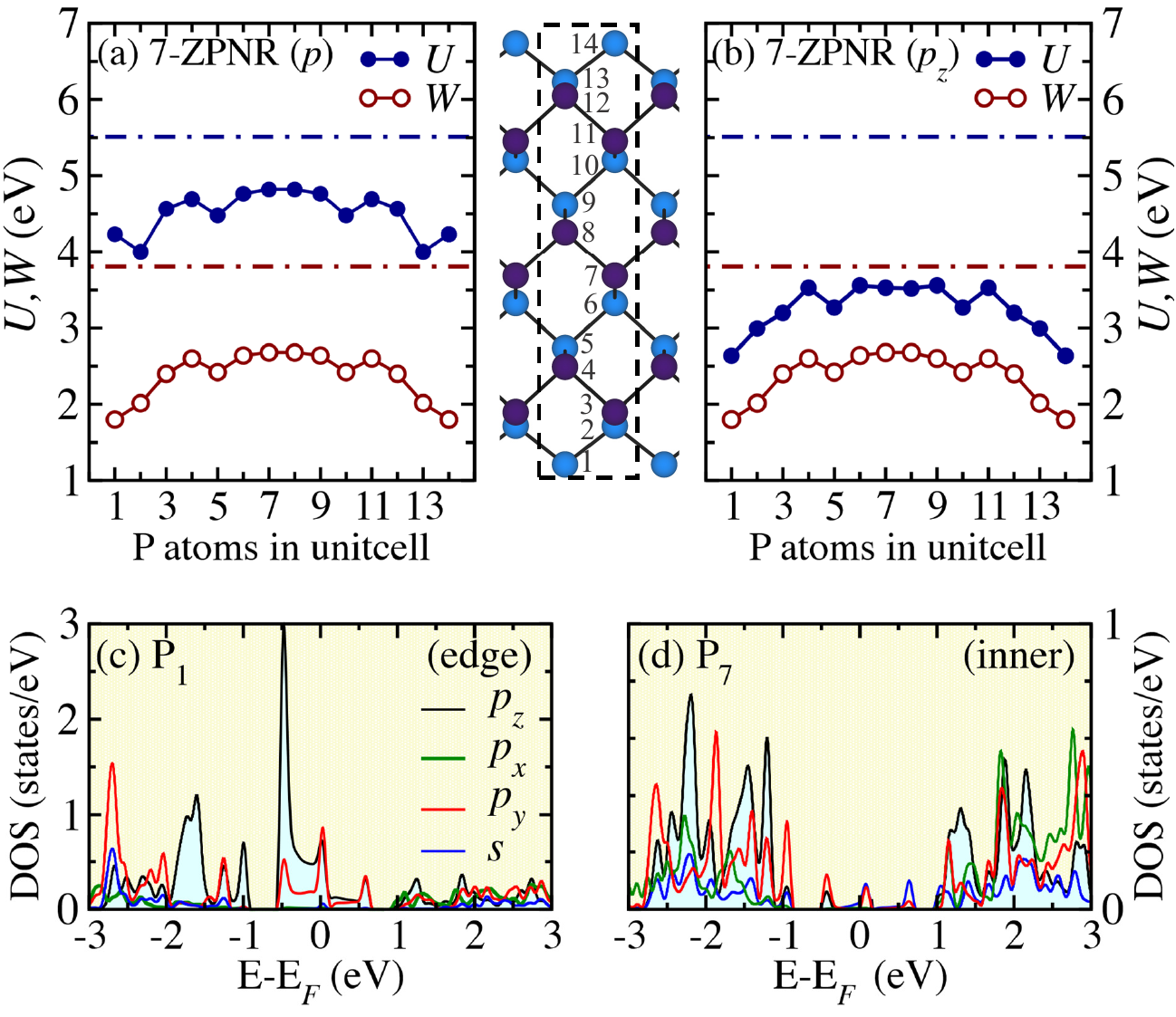}
	\vspace{-0.7 cm}
	\caption{(Color online) On-site Coulomb interaction $U$ (blue line) and fully screened Coulomb interaction $W$ (red line) of (a) $p$ and (b) $p_{z}$ selected correlated subspace for 7-ZPNR. Projected density of states for (c) edge and (d) inner P atoms of 7-ZPNR.}
	\label{fig:subm10}
\end{figure}

In this section, we discuss the Coulomb interaction of metallic bare ZPNRs. 
As shown in Fig.\,\ref{fig:subm2}(c) PNRs with unsaturated zigzag edges exhibit quasi-flat edge bands with $p_z$ character at E$_F$. 
It is important to note that the parameters used in the low-energy model Hamiltonian are defined for a system without spontaneous symmetry breaking, i.e., non-spin-polarized or paramagnetic metal. Therefore, the calculation of effective Coulomb interaction parameters should be based on the non-spin-polarized case.
In Figs.\,\ref{fig:subm10}(a) and \ref{fig:subm10}(b), we present the on-site $U$ and $W$ parameters for the $p$ and $p_z$ orbitals as the correlated subspaces of the 7-ZPNR, respectively. As shown, due to the presence of metallic edge states in the single-particle spectrum of bare ZPNRs [see Figs.\,\ref{fig:subm10}(c) and \,\ref{fig:subm10}(d)] the partially (fully) screened Coulomb interactions are significantly reduced at the edges of bare zigzag nanoribbons compared to hydrogen-passivated ZPNRs. In addition, this metalic edge states leads to a decrease in the Coulomb interaction compared to pristine black phosphorene and other semiconducting PNRs. This reduction is particularly pronounced for the $W$ parameters. The close energy levels of the $p_{z}$ and $p_{y}$ states enhance the contribution of $p_{z}$ $\rightarrow$ $p_{z}$, $p_{y}$ $\rightarrow$ $p_{y}$, and $p_{z}$ $\rightarrow$ $p_{y}$ transitions to the polarization function. As a result, the $W$ parameters are substantially reduced.

\begin{figure}[b]
	\centering
	\includegraphics[width=87mm]{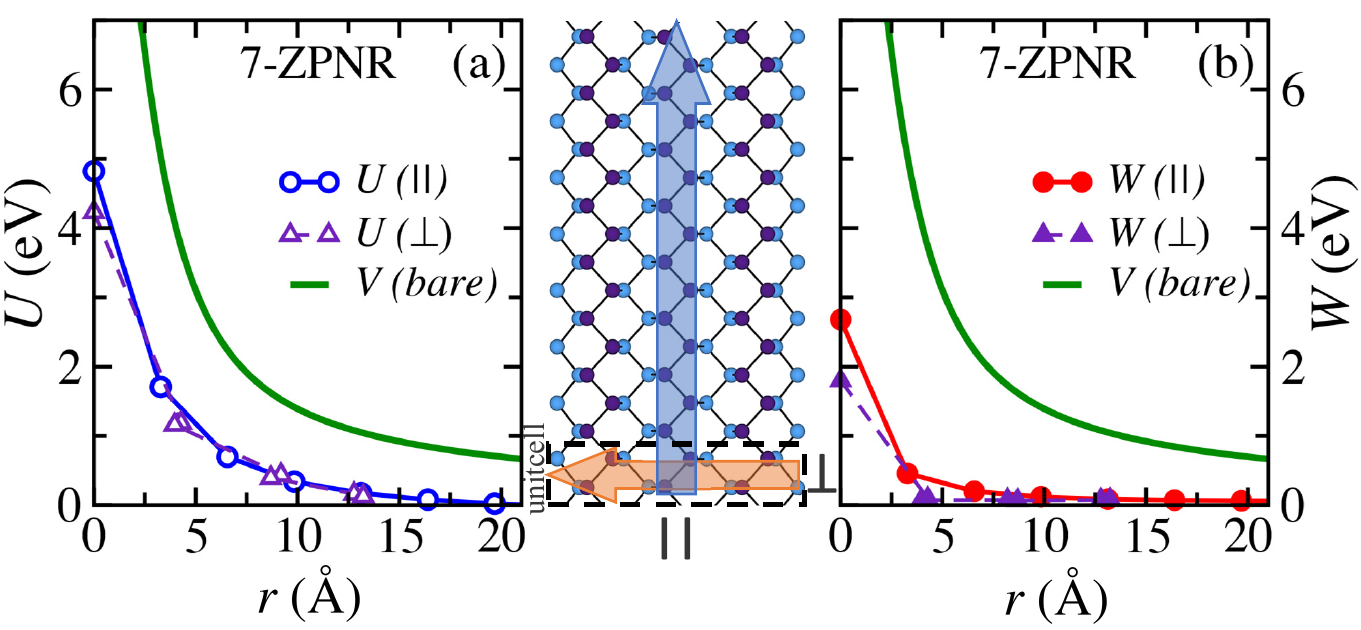}
	\vspace{-0.6 cm}
	\caption{(Color online) (a) Partially $U$ and (b) fully $W$ screened inter-site Coulomb interaction for $p$ electrons as a function of distance $r$ between two P atoms along $(\parallel)$ and acrros $(\perp)$ the ribbon for 7-ZPNR. The bare interaction $V (r)$ is presented with the green line.}
	\label{fig:subm11}
\end{figure}

The single band at E$_F$ is not purely of $p_z$ character, but it is partially mixed with $p_y$ orbitals. Therefore, terms like $p_z$ are used to describe their dominant orbital character. However, due to the great importance of the single $p_z$ band, which almost allows defining an effective single band low-energy Hamiltonian, we have also reported the electron-electron interactions for this single band $p_z$ correlated subspace.
For comparison, in Fig.\,\ref{fig:subm10}(b), the corresponding results are presented considering the $p_{z}$ orbitals as a correlated subspace in the cRPA framework. Incorporation of the efficient screening through $p_{y}$ $\rightarrow$ $p_{y}$ and $p_{z}$ $\rightarrow$ $p_{y}$ taking place in these narrow bands gives rise to smaller interaction parameters $U$ between the localized $p_z$ electrons compared to the $p$ selected correlated subspace.
To characterize the long-range behavior of the Coulomb interactions in bare ZPNRs, we present the off-site $U$ and $W$ parameters for the 7-ZPNR as a function of the distance $r$ between two P atoms in Figs.\,\ref{fig:subm11}(a) and \,\ref{fig:subm11}(b).
Due to the presence of metallic edge states, the Coulomb interaction is fully screened at relatively short distances, around 10 {\AA}. The strength of correlation is defined as the ratio of the effective Coulomb interaction $U$ to the bandwidth $W_b$. From the DFT-PBE band structure in Fig.\,\ref{fig:subm3}(c), we obtained a bandwidth of 1.2 eV of the 7-ZPNR. We find that $U/W_{b} > 1$, indicating metallic ZGNRs can be considered as a correlated system.

The large value of edge states at E$_F$ and the local Coulomb interaction leads to the instability of the paramagnetic state and tends to induce spin polarization. A stable magnetic state at the edges of ZPNR for $U>1.4t$ was reported, which $t$ is the nearest-neighbor hopping energy~\cite{Vahedi}. Aditionally,  QMC and DMFT calculations suggest that zigzag PNRs exhibit room temperature ferromagnetism at their edges, with a small energy difference per spin between the ferromagnetic (FM) and antiferromagnetic (AFM) spin-aligned ground states~\cite{Vahedi,Yang}. So, in the following, we discuss the appearance of ferromagnetism in the bare zigzag PNRs by using Stoner's model. For itinerant ferromagnetism, the instability of the paramagnetic state is given by the Stoner criterion $U$.$N$(E$_F$)$>$1. Table\,\ref{table3} presents the DOS at E$_F$ in the nonmagnetic state $N$(E$_F$), magnetic moments, and the Stoner criterion $U$.$N$(E$_F$) for all P atoms across the ribbon of 7-ZPNR. Only the P atoms located at the edges of the ribbon meet the Stoner criterion. Also, spin-polarized calculations based on spin-polarized DFT+PBE show a sizable magnetic moment 0.64$\mu_{B}$ for the edge atoms, which rapidly destroy towards the center of the ribbon. These edge magnetic moments are coupled ferromagnetically along the ribbon and antiferromagnetic from edge to edge. According to recent SQUID magnetometry and electron paramagnetic resonance (EPR) magnetization probes, the edge of PNRs displays macroscopic magnetic characteristics at room temperature, with internal fields ranging from approximately 250 to 800 mT~\cite{Ashoka}. Our results are in agreement with the QMC and DMFT prediction of two vying AFM and FM ground states in bare ZPNRs~\cite{Vahedi,Yang}.

\begin{figure}[b]
	\centering
	\includegraphics[width=75mm]{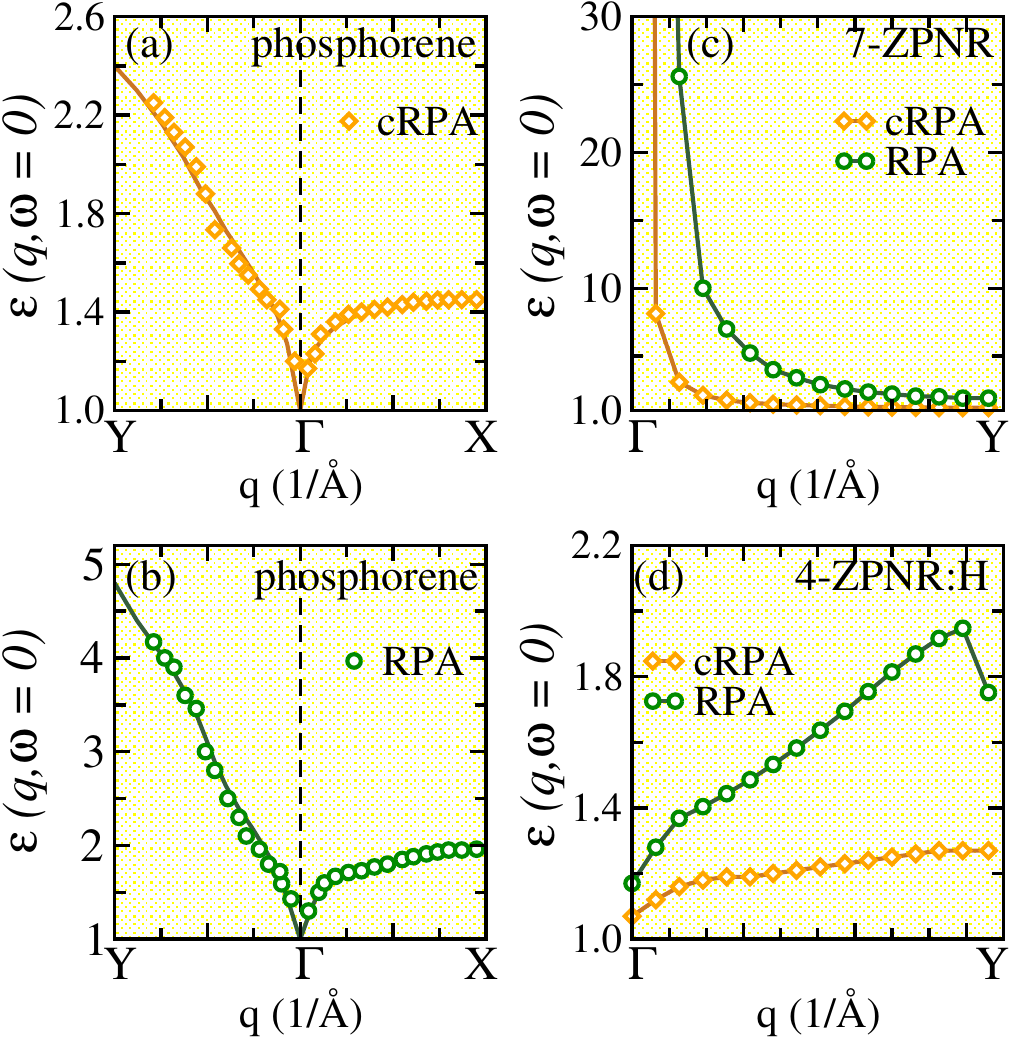}
	\vspace{-0.1 cm}
	\caption{(Color online) Static (a) cRPA and (b) RPA dielectric functions $\epsilon (q,\omega=0)$ of
		black phosphorene as a function of momentum transfer $q$ along with zigzag ($\Gamma$ $\rightarrow$ $X$) and armchair ($\Gamma$ $\rightarrow$ $Y$) directions of Brillouin zone. Static cRPA (RPA) dielectric functions $\epsilon (q,\omega=0)$ of
		(c) metallic 7-ZPNR and (d) semiconducting 4-ZPNR:H as a function of momentum transfer $q$.}
	\label{fig:subm12}
\end{figure}

In semiconducting low-dimensional systems, the macroscopic screening is nonlocal, and it is characterized by a $q$-dependent macroscopic dielectric function. In the following, we examine the variation of the static dielectric function with momentum transfer $q$ for black phosphorene, metallic 7-ZPNR, and semiconducting 4-ZPNR:H. Figs.\,\ref{fig:subm12}(a) and \,\ref{fig:subm12}(b) show static cRPA and RPA dielectric functions $\epsilon (q,\omega=0)$  calculated for monolayer phosphorene over the zigzag ($x$) and armchair ($y$) directions. 
As expected,  the value of $\epsilon (q,\omega=0)$ in $q\rightarrow0$ goes to one, and the screening becomes virtually negligible when approaching the long-wavelength limit. Overall, both cRPA and RPA  dielectric functions follow an almost identical trend.
Similar to the anisotropy of $r$-dependent inter-site interaction shown in Fig.\,\ref{fig:subm4}, the $\epsilon (q,\omega=0)$ is highly anisotropic at the edge of the Brillouin zone, which leads to the ratio of the $\epsilon (q_y)$/$\epsilon (q_x)$ $\approx$ 2.2. These findings concur with the experimental observation of highly anisotropic excitons, where the light emitted is predominantly polarized along the armchair direction~\cite{Wang,Maciej}. Indeed, the mobility of carriers in the dispersive band along the armchair direction is higher than in the flat band along the zigzag direction.

Finally, we discuss the behavior of $\epsilon (q,\omega=0)$ for 7-ZPNR and 4-ZPNR:H. As shown in Fig.\,\ref{fig:subm12}(c) for 7-ZPNR,  due to the presence of metallic edge states, the screening is strong, and the Coulomb interaction is short-range. So, at the vicinity of the $\Gamma$ point the dielectric function increases sharply with $\epsilon$($q\rightarrow0$)$\rightarrow \infty$. This is consistent with Fig.\,\ref{fig:subm11}, in which we obtain $W(r)$ drops sharply and becomes zero at short distances for 7-ZPNR. In the case of semiconducting 4-ZPNR:H [see Fig.\,\ref{fig:subm12}(d)], the situation is similar to 2D black phosphorene with the difference that the dielectric function is significantly smaller in the case of the nanoribbon.

 \begin{table}[t]
	\caption{Stoner criterion ($U.N$(E$_F$)) along with DOS at Fermi level E$_F$ ($N$(E$_F$)), Hubbard $U$, and magnetic moment (MM) of different P atoms for 7-ZPNR.}
	\centering
	\begin{tabular}{ccccccccc}
		\hline
		\hline
		\rule{0pt}{5mm}%
		P index &  & $N$(E$_F$) (1/eV)&  & $U$(eV) &  & $U.N$(E$_F$) &  & MM($\mu_{B}$)  \\ \hline
		1       &  & 1.58            &  & 4.23   &  & 6.68       &  & 0.64       \\
		2       &  & 0.06            &  & 4.00   &  & 0.24       &  & -0.01      \\
		3       &  & 0.14            &  & 4.56   &  & 0.64       &  & 0.1        \\
		4       &  & 0.03            &  & 4.69   &  & 0.14       &  & -0.02      \\
		5       &  & 0.09            &  & 4.48   &  & 0.40       &  & 0.05       \\
		6       &  & 0.04            &  & 4.76   &  & 0.19       &  & $\sim$0.00 \\
		7       &  & 0.07            &  & 4.82   &  & 0.34       &  & 0.03       \\
		8       &  & 0.07            &  & 4.82   &  & 0.34       &  & -0.03      \\
		9       &  & 0.04            &  & 4.76   &  & 0.19       &  & $\sim$0.00 \\
		10      &  & 0.08            &  & 4.47   &  & 0.36       &  & -0.05      \\
		11      &  & 0.04            &  & 4.69   &  & 0.19       &  & 0.02       \\
		12      &  & 0.13            &  & 4.56   &  & 0.59       &  & -0.1       \\
		13      &  & 0.06            &  & 4.00   &  & 0.24       &  & 0.01       \\
		14      &  & 1.57            &  & 4.23   &  & 6.64       &  & -0.64      \\ \hline
		\hline
	\end{tabular}
	\label{table3}
\end{table}

\section{conclusion}\label{sec4}

We calculated the effective on-site and off-site Coulomb interactions between $p$ electrons in single-layer of black phosphorene, bulk black phosphorene, and phosphorene 
nanoribbon by employing parameter-free cRPA scheme. These Coulomb interactions provide a fundamental
understanding of contraversal exciton excitation spectra with large exciton binding energies, the correlated phenomena such as magnetic
ordering, and Mott phase.  Also, these effective cRPA parameters can be used in model Hamiltonians thus increasing the predictive power of model calculations.
The on-site Coulomb interaction (Hubbard $U$) for black phosphorene is 2$-$3 eV less than corresponding ones in graphene. 
Due to the structural puckering, the bands with $s$, $p_x$, and $p_y$ character moves toward $p_z$ band which 
increases the contribution of $p_{x}$/$p_{y}$ $\rightarrow$ $p_{z}$ transitions to the polarization function
and reduces the Coulomb interactions. 
We have found distinctive anisotropic behavior in the long-range Coulomb interaction along the zigzag and armchair directions within phosphorene at the static limit. As a result of quantum confinement, the screening is significantly reduced in semiconducting PNRs and is heavily reliant on the ribbon width and edge passivation. Our investigation of long-range interaction has unveiled an unconventional screening phenomenon in semiconducting nanoribbons.  We have uncovered that screening actually amplifies the electron-hole interaction for separations exceeding a critical distance $r_c$, contrary to the conventional behavior observed in typical semiconductors. This unique "antiscreening" region provides insight into the significantly high experimentally extrapolated exciton binding energies of phosphorene nanoribbons.
The strength of this antiscreening effect diminishes with increasing ribbon width and edge hydrogenation. Additionally, due to the puckering structure of phosphorene, the antiscreening effect in PNRs is less pronounced than that in GNRs. In unpassivated zigzag nanoribbons, a metallic screening channel originating from quasi-flat edge bands at E$_F$ leads to $U/W_b > 1$ (where $W_b$ is the bandwidth) and a substantial gradient of inter-site Coulomb interactions, rendering them as correlated materials. We have examined the instability of the paramagnetic state of bare ZPNRs towards ferromagnetism using a Stoner model based on the calculated Hubbard $U$ parameters. Only edge P atoms can give rise to magnetic instability, which is confirmed by spin-polarized DFT calculations.

	\subsection*{Acknowledgements}
	We acknowledge Ersoy \c{S}a\c{s}{\i}o\u{g}lu and Christoph Friedrich for useful discussions.

\end{document}